# Comparing Ocean Forecasts Driven with Machine Learning-based and Physics-based Atmospheric Forcings

Xiaobing Zhou[*], Frank Colberg, Debra Hudson, Yonghong Yin,

Griffith Young, Christopher Bladwell and Catherine Deburgh-Day

Australian Bureau of Meteorology

700 Collins Street, Docklands VIC 3008, Australia

Corresponding author: xiaobing.zhou@bom.gov.au

# Abstract

Operational ocean forecasting systems conventionally employ dynamical ocean models driven by atmospheric forcing derived from numerical weather prediction (NWP) models. Recent advancements in artificial intelligence and machine learning (ML) have led to the development of ML-based atmospheric weather models, which have competitive, if not better, medium range forecast accuracy compared to traditional NWP systems. This study evaluates the impact of ML-based atmospheric forcing on ocean forecast skill through two sets of 10-day forecasts using the UK Met Office GOSI9 configuration of the NEMO dynamical ocean model. Both experiments share identical ocean initial conditions; but differ in atmospheric forcing: one uses ECMWF's ML-based AIFS model, while the other uses the Australian Bureau of Meteorology's physics-based NWP model, ACCESS-G3. Forecasts were initialized on the first day of each month over the period 2023–2024. The quality of the atmospheric forcing was assessed by comparing AIFS and ACCESS-G3 forecast skill against both ECMWF reanalysis v5 (ERA5) and ACCESS-G3 analyses. Results indicate that AIFS consistently outperforms ACCESS-G3, either from the initial forecast time or after the first few days. Oceanic forecast skill was evaluated against both the GOSI9 reanalysis and observations, focusing on key surface variables including sea surface temperature, salinity, sea level, and ocean currents. The ocean forecasts forced with AIFS atmospheric data exhibit comparable or enhanced predictive skill compared to those forced with ACCESS-G3 data. These findings underscore the potential of ML-based atmospheric models to replace traditional NWP forcing in operational ocean forecasting systems, offering improved accuracy and computational efficiency.

## 1. Introduction

Accurate ocean forecasts are essential for delivering routine oceanographic information that supports a wide range of decision-making processes. These include marine safety and navigation (e.g., Chan et al., 2022), fisheries management (e.g., Hilborn et al., 2020), environmental monitoring and conservation (e.g., Halpern et al., 2015), disaster preparedness and response (e.g., Bernard and Titov, 2015), policy development (e.g., Winther et al., 2020), and the planning of renewable energy initiatives (e.g., Gatzert and Kosub, 2016).

Physics-based numerical systems are widely used to forecast key ocean variables such as temperature, salinity, sea surface height, and currents. These forecasts underpin predictions of ocean phenomena such as storm surges, mesoscale eddies, fronts, coastally trapped waves, coastal upwelling, and marine heatwaves. Examples of operational ocean forecast systems include OceanMAPS (Australian Bureau of Meteorology; Brassington et al. 2023), the Mercator Ocean System (France/CMEMS; Lellouch et al., 2018), FOAM (UK Met Office; Mignac et al. 2025) and HYCOM/GOFS (US Navy/NOAA; Cummings and Smedstad 2013). These ocean models rely on atmospheric forcing from traditional numerical weather prediction (NWP) models. For instance, OceanMAPS (Brassington et al. 2023) uses atmospheric forcing from the Bureau of Meteorology's global NWP model, ACCESS-G.

Recent advances in machine learning (ML) are transforming weather prediction, leading to the development of ML-based weather models, such as FourCastNet (Pathak et al., 2022), Pangu-Weather (Bi et al., 2023), GraphCast (Lam et al, 2023), FuXi (Chen et al.,2023), AIFS (Artificial Intelligence Forecasting System; Lang et al., 2024a), GenCast (Price et al., 2025) and Aurora (Bodnar et al., 2025). These models have demonstrated comparable or superior accuracy to traditional NWP systems across various variables and lead times, while offering significant gains in computational speed.

In parallel, ML-based models for global, medium-range, eddy-resolving ocean forecasting are beginning to emerge (e.g., Wang et al., 2024; Aouni et al., 2025; Jia et al 2025), but their development is progressing far more slowly than that of ML-based weather models (Aoni et al. 2025; Jia et al. 2025). This disparity raises an important question: can ML-based weather models provide sufficiently accurate atmospheric forcing to drive dynamical ocean forecasts? To our knowledge, this question has not yet been addressed in the literature.

In this study, we investigate the feasibility of using ML-based atmospheric forcing to drive physics-based ocean forecasts. Specifically, we compare the forecast performance of a dynamical ocean model, the UK Met Office's latest Global Ocean and Sea Ice model (GOSI configuration version 9, Guiavarc'h et al. 2025), forced with atmospheric forecasts from ECMWF's AIFS v1.0 ML weather forecast model (Lang et al., 2024a; Moldovan et al., 2025) and the Australian Bureau of Meteorology's ACCESS-G3 physics-based global weather forecast model (Bureau of Meteorology National Operational Centre, 2019). By using identical ocean initial conditions, we isolate the impact of the atmospheric forcing and assess whether ML-based weather models have the potential to support operational ocean forecasting. We also force the ocean model with ERA5 reanalysis data (Hersbach et al., 2020) to provide an upper limit benchmark for ocean forecast performance using "observed" atmospheric forcing.

AIFS v1.0 (also referred to as AIFS Single) is ECMWF's first fully operational ML-only forecasting system, implemented in February 2025 (Lang et al., 2024a; Moldovan et al., 2025). It produces medium-range weather forecasts that are competitive with their traditional Integrated Forecasting System (IFS). The model is based on a graph neural network (GNN) encoder–decoder and a sliding-window transformer processor, operating on a 0.25° global grid with 13 pressure levels. AIFS was trained on ERA5 reanalysis data spanning 1979–2022 and then fine-tuned on ECMWF operational analysis from 2016-2022.

ACCESS-G3 is the third-generation global NWP model developed by the Australian Bureau of Meteorology. Built on the UK Met Office's Unified Model framework, ACCESS-G3 runs four times daily (00Z, 06Z, 12Z, and 18Z) and produces forecasts up to 240 hours. Its initialization employs a hybrid 4D-Var data assimilation scheme with a 6-hour window, ensuring that the model state reflects the most recent observations.

This study is organised as follows. Section 2 will describe the models, experimental design and verification methods. Section 3 will evaluate weather forecasts from AIFS and ACCESS-G3, to compare the quality of the atmospheric forcing provided to the ocean model. Section 4 will present the results of the ocean forecast experiments, evaluating the impact of the atmospheric forcing on forecast skill for sea surface temperature (SST), sea surface height (SSH), sea surface salinity (SSS), zonal sea surface current (SSU) and meridional sea surface current (SSV). The final section will provide a summary and discussion.

## 2. Ocean model, experimental design and verification methods

### 2.1 Ocean model description

The UK Global Ocean and Sea Ice configuration version 9 (GOSI9) (Guiavarc'h et al. 2025) from the Met Office is used in this study. The Bureau of Meteorology is currently building its next operational ocean forecasting system using this model configuration.

GOSI9 is built on NEMO 4.0.4 (Madec and NEMO system team, 2019) and uses the $SI^3$ sea ice component (Blockley et al., 2024). The model grid is eORCA025 with a horizontal resolution of ¼°. This grid has 1442 points in longitude and 1207 in latitude, with grid distances ranging from 28 km at the Equator to 6 km at high latitudes. The quasi-isotropic bipolar grid in the Northern Hemisphere avoids a singularity point in the Arctic Ocean. There are 75 vertical levels, ranging in thickness from 1 meter near the surface to 200 meters at 6,000 meters. The high resolution near the surface is capable of resolving the diurnal cycle (Bernie et al., 2005), the second baroclinic mode and the surface layer (Stewart et al., 2017).

## 2.2 Experimental design and atmospheric forcing

Three ocean forecast experiments are conducted in this study. All experiments share identical oceanic initial conditions and cover the same prediction period, differing only in the atmospheric forcing applied, namely: 1) AIFS forecast forcing data, 2) ACCESS-G3 forecast forcing data, and 3) ERA5 reanalysis data. The ERA5-based experiment is treated as the reference run, representing the upper benchmark for ocean prediction skill obtained with "observed" atmospheric forcing.

The AIFS model provides 6-hourly instantaneous fields for air temperature and wind, along with accumulated values for precipitation, snowfall, and surface downward shortwave and longwave radiation fluxes. The model operates at a horizontal resolution of 0.25°. In this study, ERA5 data are used to initialize the AIFS forecasts. ERA5 has the same horizontal resolution as AIFS outputs but offers data at a 1-hour temporal frequency. In contrast, the ACCESS-G3 forecast system provides hourly mean values for all variables, with accumulated values for precipitation and snowfall. It features a higher horizontal resolution than AIFS, with grid spacing of approximately 12 km in mid-latitudes and 17 km in tropical regions.

The prediction period for the experiments spans 2023–2024. This timeframe is selected because the AIFS model was pre-trained on the ERA5 dataset covering the years 1979–2022 (Moldovan et al, 2025; https://confluence.ecmwf.int/display/FCST/Implementation+of+AIFS+Single+v1 ). Since 2023–2024 lies outside the AIFS training period, it provides an opportunity for a fair and unbiased comparison between the forecasts produced by AIFS and those from the ACCESS-G3 model. Forecasts are initialized on the first day of each month, with a forecast range of 10 days. This results in a total of 24 forecast cases for each experiment.

## 2.3 Ocean initial conditions

The ocean initial conditions for the forecasts are obtained from a data assimilation (DA) system based on the GOSI9 model. The ocean model is forced with ERA5 atmospheric reanalysis data provided by the European Centre for Medium-Range Weather Forecasts (ECMWF). The DA system employs an ensemble Kalman filter operated in a hybrid mode, comprising 48 dynamic and 144 static ensemble members (Sakov, 2014). Assimilated oceanic observations include SST data from the Visible Infrared Imaging Radiometer Suite on the Suomi-NPP satellite and NASA's Advanced Microwave Scanning Radiometer 2; along-track altimeter sea level anomaly data from the Radar Altimeter Database System (Scharroo et al., 2013); in situ temperature and salinity profiles from the EN4 dataset (Ingleby and Huddleston, 2007); and satellite-derived L3C sea ice concentration data from the EUMETSAT Ocean and Sea Ice Satellite Application Facility (OSI SAF; https://osi-saf.eumetsat.int/products/sea-ice-products). The data assimilation cycle length is 24 hours. The observations are assimilated asynchronously at their time of acquisition (Sakov et al., 2014). SST observations are assimilated in 2-hourly batches while daily batch processing is applied to temperature, salinity and satellite altimetry observations. The DA system has been run over the period from 2021 to early 2025.

## 2.4 Verification data and methods

The quality of the atmospheric forecast forcing data from AIFS and ACCESS-G3 are evaluated by computing the correlation and root mean square error (RMSE) between predicted daily mean values and corresponding atmospheric reanalysis data for each forecast day across the 24 cases. Only variables that are used to force the ocean model are assessed here. The GOSI9 ocean model employs the bulk formulae developed by Large and Yeager (2009) to compute turbulent flux transfer coefficients. These formulations require several forcing variables, including wind velocity, air temperature, precipitation rate, snowfall rate, specific humidity, and downward longwave and shortwave radiation fluxes. These atmospheric forcing variables play a crucial role in driving ocean forecasts. Forecast skill is computed with reference to both the ERA5 reanalysis and the ACCESS-G3 analysis to ensure robustness.

To assess the performance of the ocean forecasts, this study focuses on the prediction skill of sea surface variables, including SST, SSH, SSS, SSU and SSV. Subsurface ocean variables are not evaluated due to the short forecast lead time considered in this analysis.

Observed SST and SSH data have been assimilated into the model to generate initial conditions for the ocean forecasts. However, observed SSS and surface current data are not assimilated. Consequently, the prediction skill of SST and SSH is assessed using observed data only, while the performance of SSS, SSU, and SSV is evaluated against both ocean reanalysis and observational datasets.

The observed SST used for evaluation is the 1/4° daily optimum interpolation sea surface temperature (OISST) data (Reynolds et al. 2007). The altimeter data are global ocean gridded L4 sea surface heights (https://doi.org/10.48670/moi-00145). The observed total velocity fields (https://doi.org/10.48670/mds-00327) are obtained by combining CMEMS satellite Geostrophic surface currents and modelled Ekman currents at the surface and 15m depth using ERA5 wind stress (Rio et al. 2014). The observed daily global Level-4 SSS data at 1/8° of resolution (https://doi.org/10.48670/moi-00051), obtained through a multivariate optimal interpolation algorithm that combines sea surface salinity images from multiple satellite sources as NASA's Soil Moisture Active Passive (SMAP) and ESA's Soil Moisture Ocean

Salinity (SMOS) satellites with in situ salinity measurements and satellite SST information (Droghei et al. 2016).

## 3. Analysis of the atmospheric forcings

### 3.1 Global patterns of correlation and RMSE at day-10 of the forecast

This section examines the day 10 forecast performance for key atmospheric variables used to drive the ocean model, highlighting their spatial patterns. The next section places these results in context by assessing forecast skill across lead times from day 1 to day 10. Fig. 1 shows the forecast correlation skill at day 10 from the AIFS model (left column) and ACCESS-G3 (middle column) against ERA5 reanalysis, along with their difference in correlation (right column), for 2m-air temperature, downward shortwave radiation flux, mean sea level pressure (MSLP), and 10m-zonal wind. Correlation and RMSE results—benchmarked against ERA5 (days 1 and 5) and ACCESS-G3 analysis data (days 1, 5, and 10)—are provided in the Supplementary Material (Figs. S1–S10).

The spatial patterns of forecast skill for each variable are generally consistent between the AIFS and ACCESS-G3 models. For 2m-air temperature, both models exhibit high correlation skill across most regions, with notable exceptions in the tropics and the Southern Ocean. The reduced correlation near the equator, particularly in the Intertropical Convergence Zone (ITCZ) and South Pacific Convergence Zone (SPCZ), is likely associated with persistent cloud cover and strong convective activity, both of which are challenging to predict. The Southern Ocean presents additional forecasting challenges due to its dynamic complexity, including strong westerly winds, frequent storm systems, rapidly evolving low-pressure systems, active air–sea interactions, and sharp oceanic fronts (Rao et al., 2024; Behrens and Bostock, 2023). These factors contribute to high atmospheric variability, making accurate prediction more difficult. Overall, the AIFS model shows higher correlation skill across most regions, particularly in the lower-skill areas of the western Pacific and tropical Indian Ocean.

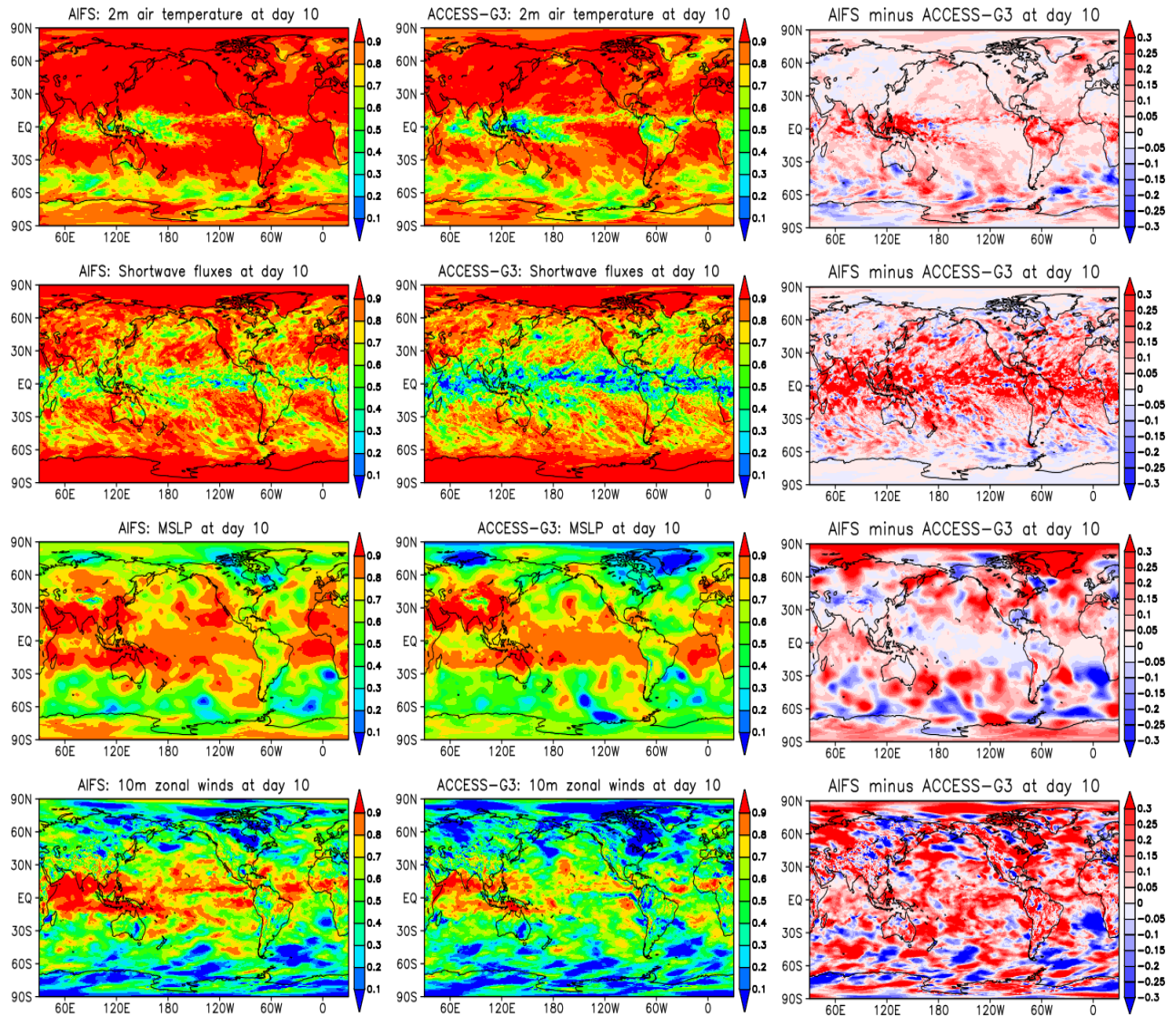

**Fig. 1.** Correlations between AIFS model forecasts and ERA5 reanalysis (left column), ACCESS-G3 forecast and ERA5 reanalysis (middle column) and the correlation differences between AIFS and ACCESS-G3 (right column) for day-10 of the forecasts. The atmospheric variables, from top to bottom, are 2m air temperature, surface downward shortwave flux, mean sea level pressure (MSLP) and 10m zonal wind.

For downward shortwave radiation fluxes, the lowest forecast skill is observed in tropical regions, coinciding with areas of deep convective cloud activity. In these zones, shortwave radiation is strongly influenced by rapidly evolving cloud systems, which pose significant challenges for prediction. In contrast, higher skill is found in the dry zones of the mid- and high latitudes, where atmospheric conditions are more stable and less cloud-covered, making them easier to forecast. Compared to ACCESS-G3, the AIFS model exhibits higher correlation skill across most regions, particularly in the low and mid-latitudes.

There is significantly higher forecast skill for MSLP compared to surface downward shortwave radiation at low and subtropical latitudes. This is consistent with MSLP being a more slowly varying, large-scale field governed by synoptic and planetary-scale dynamics. The lowest correlation values are typically found in regions characterised by high transient variability, such as the Southern Ocean storm tracks and polar areas. Both models maintain strong predictive skill for MSLP even at a 10-day lead time, underscoring the inherent predictability of large-scale pressure systems. Compared to ACCESS-G3, the AIFS model exhibits higher correlation skill in the tropical Indian Ocean and in the middle and high-latitudes.

For surface zonal winds, the highest forecast skill is observed along the equatorial belt, particularly within the Indo-Pacific warm pool. In this region, zonal wind variability is strongly influenced by large-scale, slowly evolving phenomena such as the Madden–Julian Oscillation (MJO) and the El Niño–Southern Oscillation (ENSO), which enhance predictability. In contrast, the lowest skill is found in high-latitude regions, especially over the Southern Ocean and the Arctic, where atmospheric flow is dominated by transient, synoptic-scale variability. These rapidly changing conditions limit forecast accuracy beyond a one-week lead time. In general, AIFS model shows higher correlation skill in majority of areas.

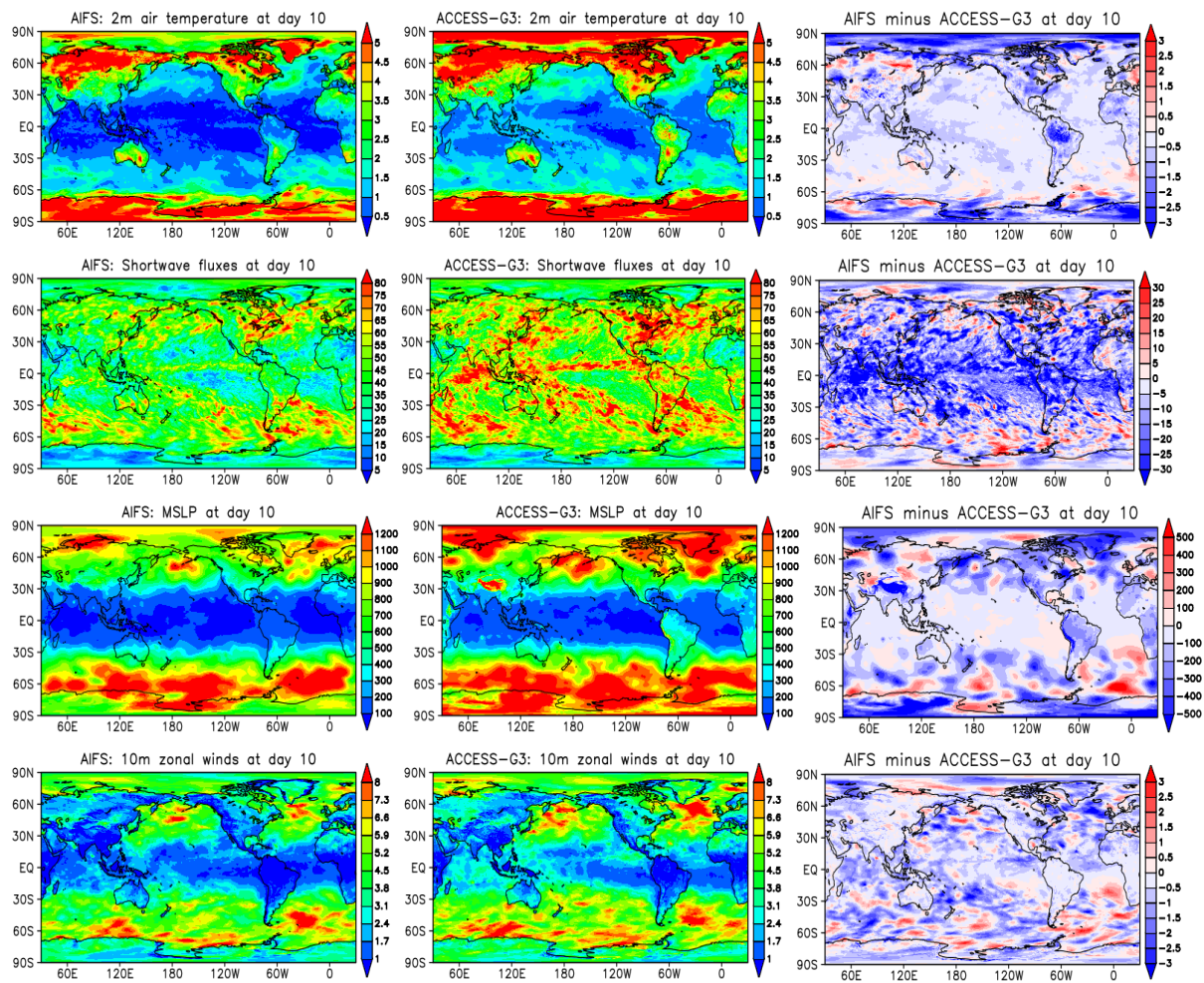

**Fig. 2.** Same as Fig. 1 except for root mean square error (RMSE). The units are K for air temperature, W/m$^2$ for short wave fluxes, Pa for MSLP and m/s for winds.

Fig. 2 presents the same set of variables as Fig. 1 but evaluated using RMSE instead of correlation. It is important to note that low RMSE does not always correspond to high correlation skill. While correlation assesses the phase agreement between model forecasts and reference data, RMSE is an absolute error metric that captures both model bias and variability error. Therefore, RMSE provides complementary insight into forecast performance, particularly in regions where systematic biases may exist despite good temporal alignment.

For 2m air temperature, both AIFS and ACCESS-G3 models exhibit low RMSE values in tropical regions, despite showing low correlation skill there. This suggests that while the models capture the general magnitude of temperature well, they struggle with the timing or phase of variability in these areas. In addition, the relative magnitude of the temperature fluctuations is smaller in lower latitudes. In the Southern Ocean, large RMSE values are consistent with low correlation, reflecting the challenges of forecasting in this dynamically complex region and the greater variability of temperature. In the mid-latitudes and polar regions of the Northern Hemisphere, both models show relatively high correlation skill but also large RMSE values, indicating that while the temporal patterns are well captured, systematic biases or amplitude errors may be present. Notably, the AIFS model demonstrates smaller RMSE in high-latitude regions compared to ACCESS-G3, although both models show similar RMSE magnitudes at lower latitudes.

For surface downward shortwave flux, the RMSE values in ACCESS-G3 are notably high over the Maritime Continent, ITCZ, SPCZ, the Kuroshio and Gulf Stream regions, and the Southern Ocean. These elevated RMSEs may be linked to deficiencies in the cloud representation within the ACCESS-G3 model. The RMSE values in these regions are significantly larger than those observed in the AIFS model.

In the tropics and sub-tropics, MSLP and 10-m zonal winds exhibit relatively low RMSE and high correlation skill for both models. In contrast, the mid-latitudes and polar regions show reduced correlation and increased RMSE, particularly over the Southern Ocean and near-polar areas.

For MSLP, both the AIFS and ACCESS-G3 models demonstrate comparable RMSE skill in the tropics. However, the AIFS model exhibits generally lower RMSE values at high latitudes. In the case of 10m zonal winds, the differences between AIFS and ACCESS-G3 are characterised by small-scale spatial patterns, although AIFS generally exhibits smaller RMSEs.

Additionally, 10m meridional winds and precipitation have been evaluated against ERA5 and ACCESS-G3 analysis data (not shown). The spatial distributions of correlation and RMSE skill for the surface meridional winds resemble those of surface zonal winds. For precipitation, the global patterns of correlation and RMSE are broadly similar between the AIFS and ACCESS-G3 models. When evaluated against ERA5 data at forecast day 10, the AIFS model shows moderate-to-high correlation in the tropics, particularly along the ITCZ and regions influenced by MJO. In contrast, forecast skill is weaker in the mid-latitudes and the Southern Ocean. RMSE values are highest (around 25 mm day$^{-1}$) in convectively active regions such as the tropical western Pacific and Indian Ocean, indicating challenges in capturing rainfall intensity despite reasonable skill in representing variability. In the extratropics, lower correlation reflects the influence of chaotic weather systems, while smaller RMSE values are consistent with weaker precipitation climatology. Overall, both AIFS and ACCESS-G3 models can capture large-scale precipitation patterns at medium-range lead times, but biases in rainfall intensity remain a primary source of error, particularly in tropical regions. AIFS model is generally better than ACCESS-G3 in rainfall prediction.

### 3.2 Lead-time dependent global mean correlation and RMSE skill

Figures 3 and 4 present the global mean correlation and RMSE from forecast days 1 to 10 for 2m air temperature, surface downward shortwave radiation, MSLP, and 10m zonal winds. Forecast skill is evaluated against both the ERA5 reanalysis and the ACCESS-G3 analysis. In general, higher correlation skill corresponds to lower RMSE, indicating consistent model performance across metrics.

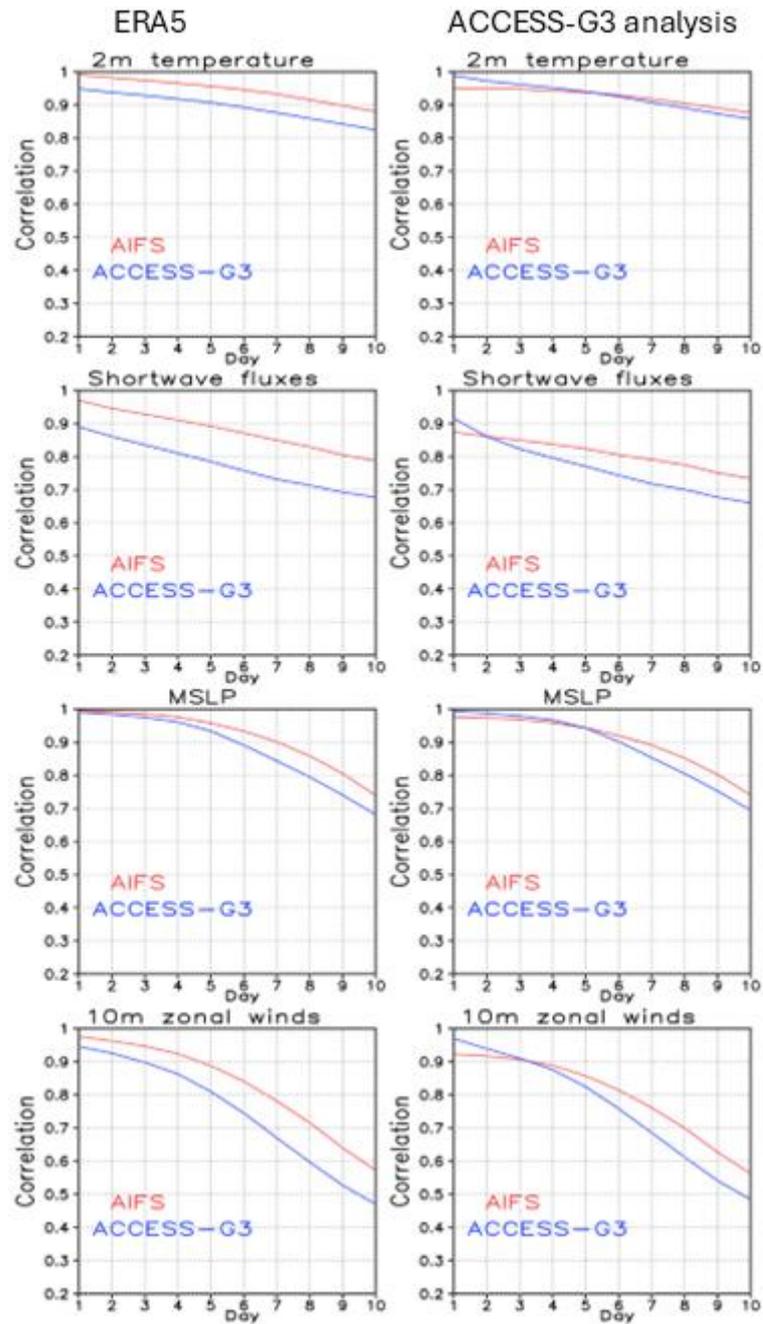

**Fig. 3.** Global mean correlations of AIFS (red line) and ACCESS-G3 (blue line) evaluated against ERA5 reanalysis (left column) and ACCESS-G3 analysis (right column) from day 1 to day 10. The atmospheric variables, from top to bottom, are 2m air temperature, surface downward shortwave flux, mean sea level pressure (MSLP), and 10m zonal wind.

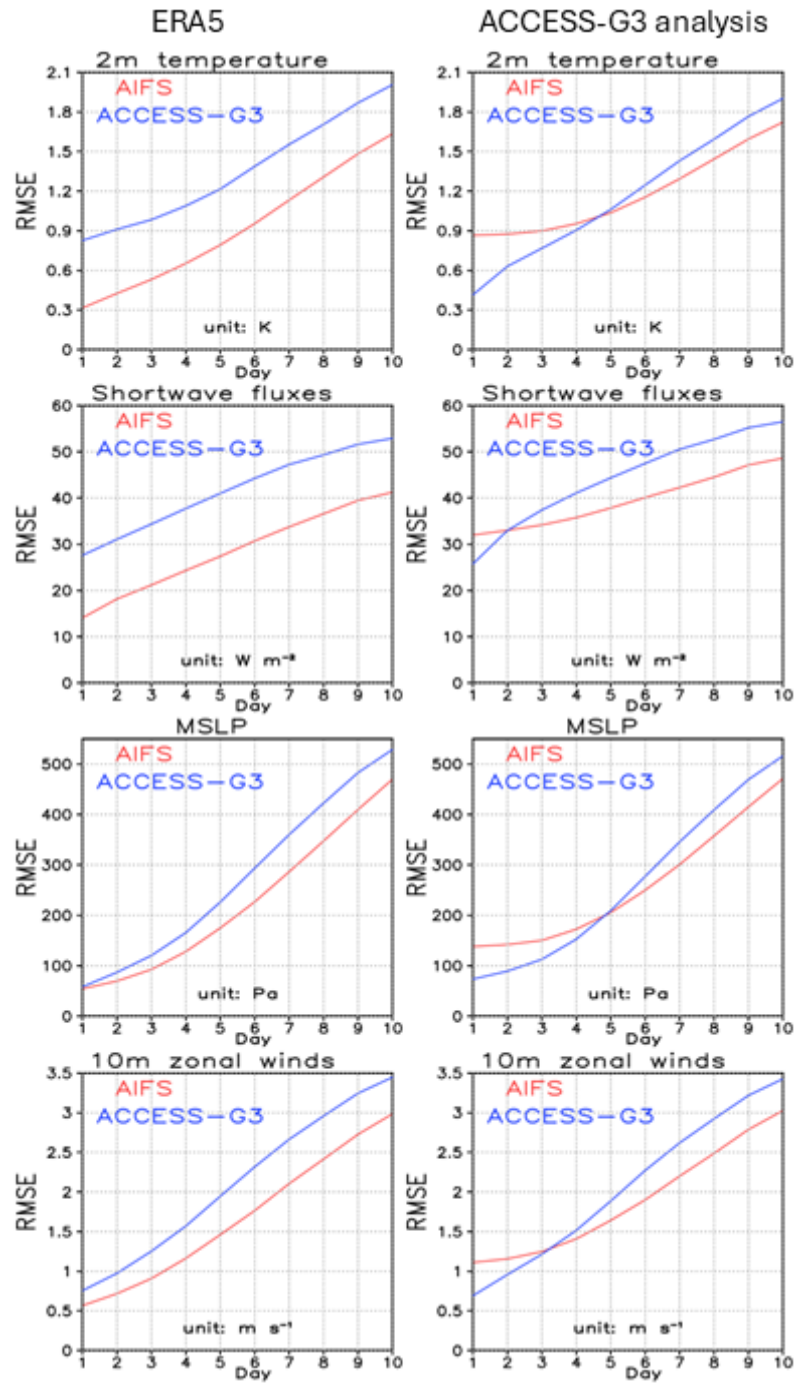

**Fig. 4.** Same as Fig. 3 except for root mean square error (RMSE).

When validated against ERA5, the AIFS model demonstrates superior skill compared to ACCESS-G3, with higher correlation and lower RMSE throughout the 10-day forecast period. However, when both models are evaluated against the ACCESS-G3 analysis, ACCESS-G3 shows better performance during the initial forecast days, but its skill declines more rapidly than AIFS thereafter. Notably, each model tends to perform better when evaluated against its own initialisation dataset, particularly at shorter lead times. In addition, for the evaluation against ERA5, AIFS has an advantage over ACCESS-G3 since it was trained on ERA5 data. Despite this, AIFS maintains higher predictive skill over longer lead times than ACCESS-G3 when assessed against both (re)analyses.

The results of Section 3 show that AIFS and ACCESS-G3 exhibit broadly similar large-scale patterns and magnitudes of skill for key atmospheric variables required to force ocean models, with AIFS often achieving higher skill, particularly at longer lead times. This indicates that AIFS has strong potential to provide reliable atmospheric forcing for dynamical ocean forecasts. In the next section, the impact of the atmospheric forcing on the performance of ocean forecasts is evaluated.

## 4. Ocean forecasts results

### 4.1 Global patterns of correlation and RMSE at day-10 of the forecast

Figures 5–8 present global maps of forecast correlation skill at day 10 for SST, SSH, SSU, SSV, and SSS, along with the correlation differences arising from the use of different atmospheric forcing datasets. The corresponding RMSE results for day 10 are shown in Figures 9–12, while correlation and RMSE results for days 1 and 5 are also provided in the Supplementary Material (Figs. S11–S26). Overall, the correlation patterns for each ocean variable at day 10 are broadly similar when the ocean model is forced with AIFS or ACCESS-G3 atmospheric data.

SST forecasts exhibit high correlation skill with observed SST data OISST across most regions (Fig. 5), with notable exceptions in the Arctic, western Pacific, tropical Indian Ocean, and Southern Ocean. The particularly low correlation in the Arctic is attributed to a lack of observations and the influence of sea ice, which dominates SST variability in seasonally or permanently ice-covered areas, reducing the impact of ocean–atmosphere fluxes. The relatively low correlation in the western Pacific and tropical Indian Ocean aligns with the skill of surface air temperature forecasts (Fig. 1) and is partly due to uncertainties in reanalysis SST data, which are affected by persistent cloud cover that limits satellite observations. These regions also show reduced inter-dataset agreement among SST reanalyses such as OISST, the operational sea surface temperature and ice analysis (OSTIA) (Good et al., 2020), and global Australian multi-sensor SST analysis (GAMSSA) (not shown). SST forecasts driven by AIFS data demonstrate slightly higher skill in the western Pacific, tropical Indian Ocean, and Arctic compared to those forced with ACCESS-G3, while performance in other regions is comparable.

At a 10-day lead time, SSH forecast skill is highest in the tropics and subtropical gyres, and markedly lower in high-latitude regions (Fig. 5). In the tropics, SSH anomalies are primarily governed by large-scale, slowly evolving equatorial waves with extended memory, contributing to greater predictability (Polkova et al., 2015). In contrast, SSH variability in mid-to-high latitudes is dominated by mesoscale eddies and baroclinic instabilities, which evolve more rapidly and thereby limit forecast skill (Chelton et al., 2011). SSH correlation patterns are similar between AIFS- and ACCESS-G3-driven forecasts, although AIFS-forced simulations show slightly higher skill in the Southern Ocean and high-latitude Northern Hemisphere.

Figures 6–8 display the day-10 correlation maps between model-forecasted SSU, SSV and SSS and the GOSI9 reanalysis, as well as with respect to independent observations. It is important to note that, unlike SST and SSH, no observed ocean current or SSS data were assimilated during model initialization in this study. As a result, the correlations between model

forecasts and the GOSI9 reanalysis are substantially higher than those obtained through validation with independent observations.

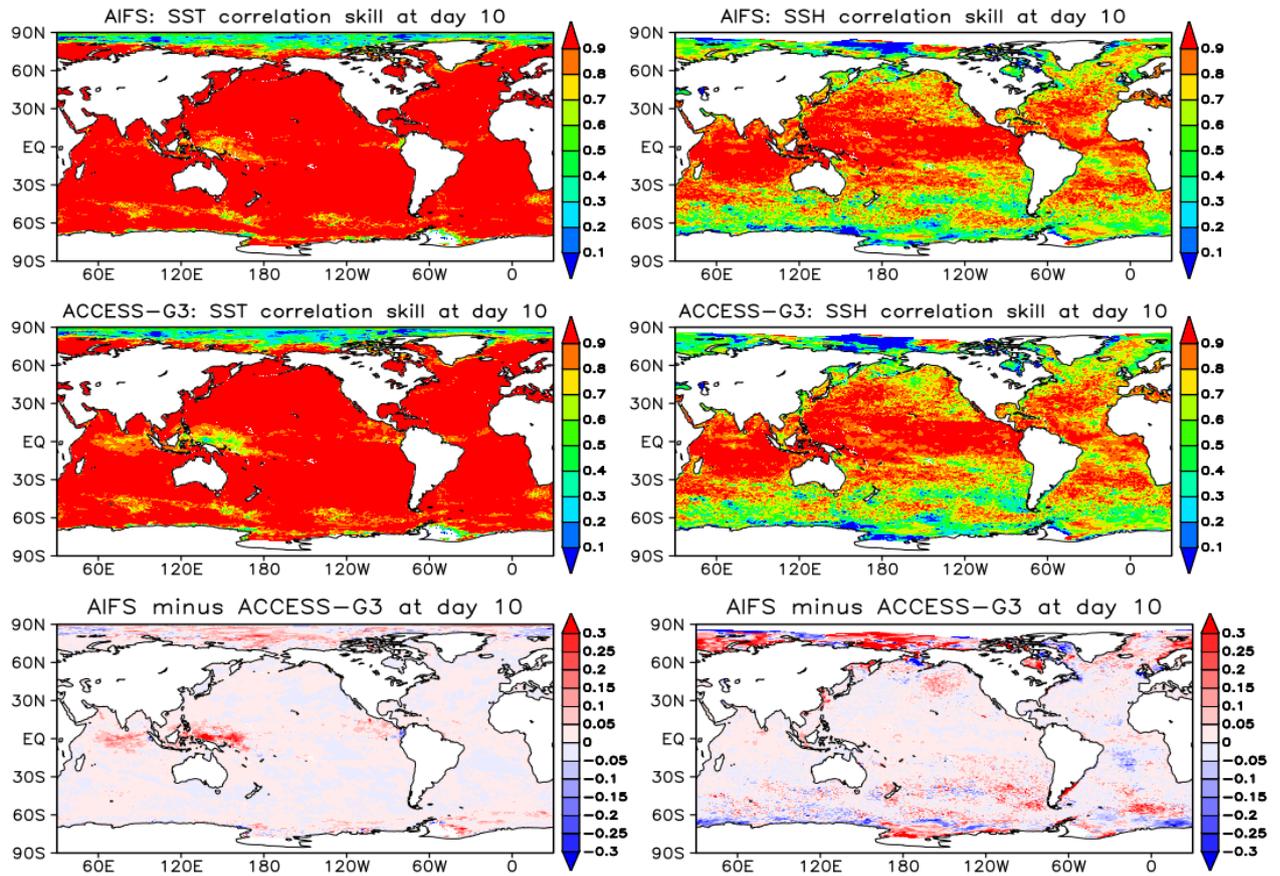

**Fig. 5.** Correlations between ocean forecasts and observations at day 10 for SST (left column) and SSH (right column). The top row shows forecasts forced with AIFS data, the middle row shows forecasts forced with ACCESS-G3 data and the bottom row shows the correlation differences between the two.

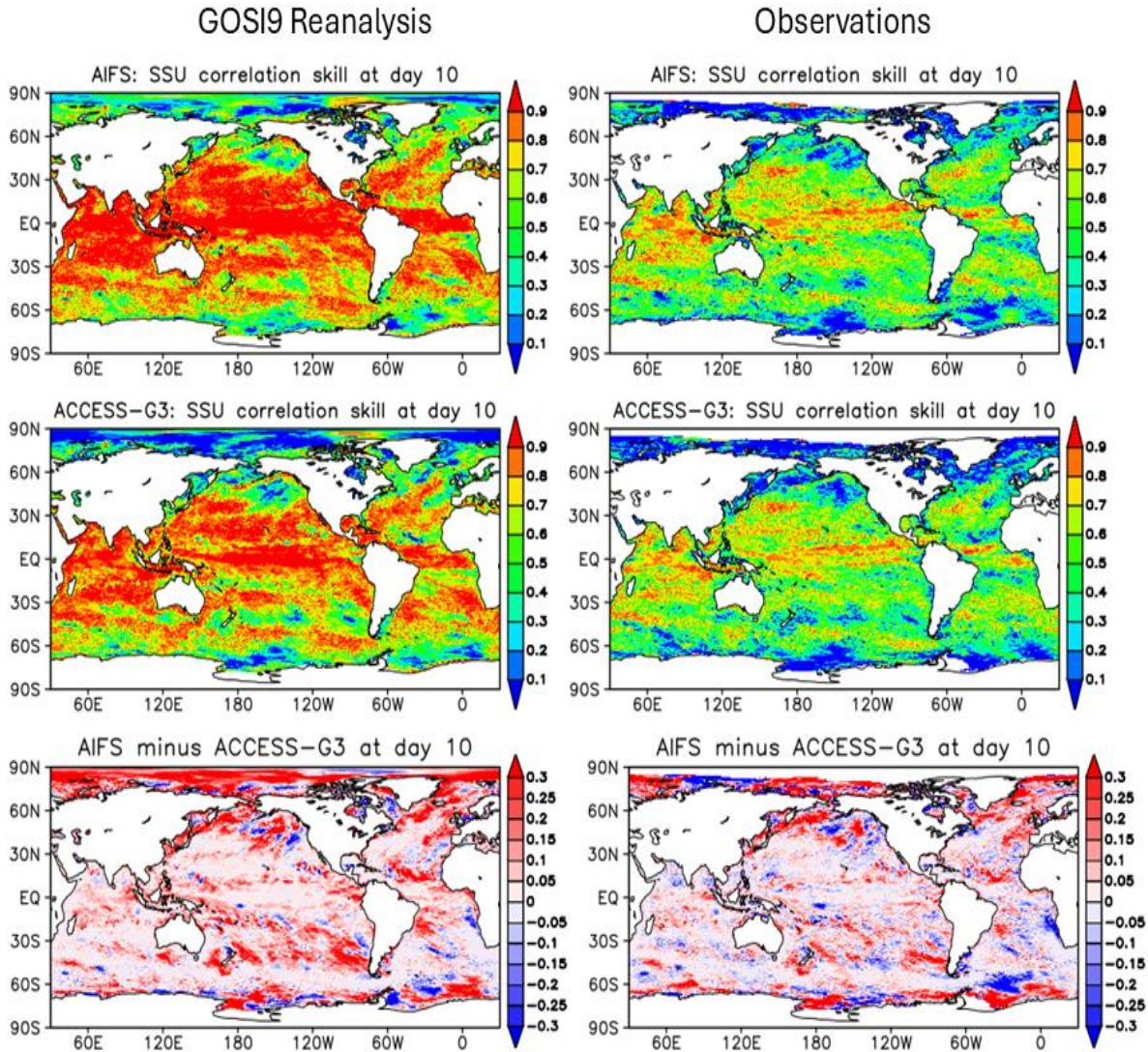

**Fig. 6.** Correlations at day 10 between model-forecasted sea surface zonal current (SSU) and GOSI9 ocean reanalysis (left column), and between model SSU and observations (right column). The top row shows forecasts forced with AIFS data, the middle row shows forecasts forced with ACCESS-G3 data and the bottom row shows the correlation differences between the two.

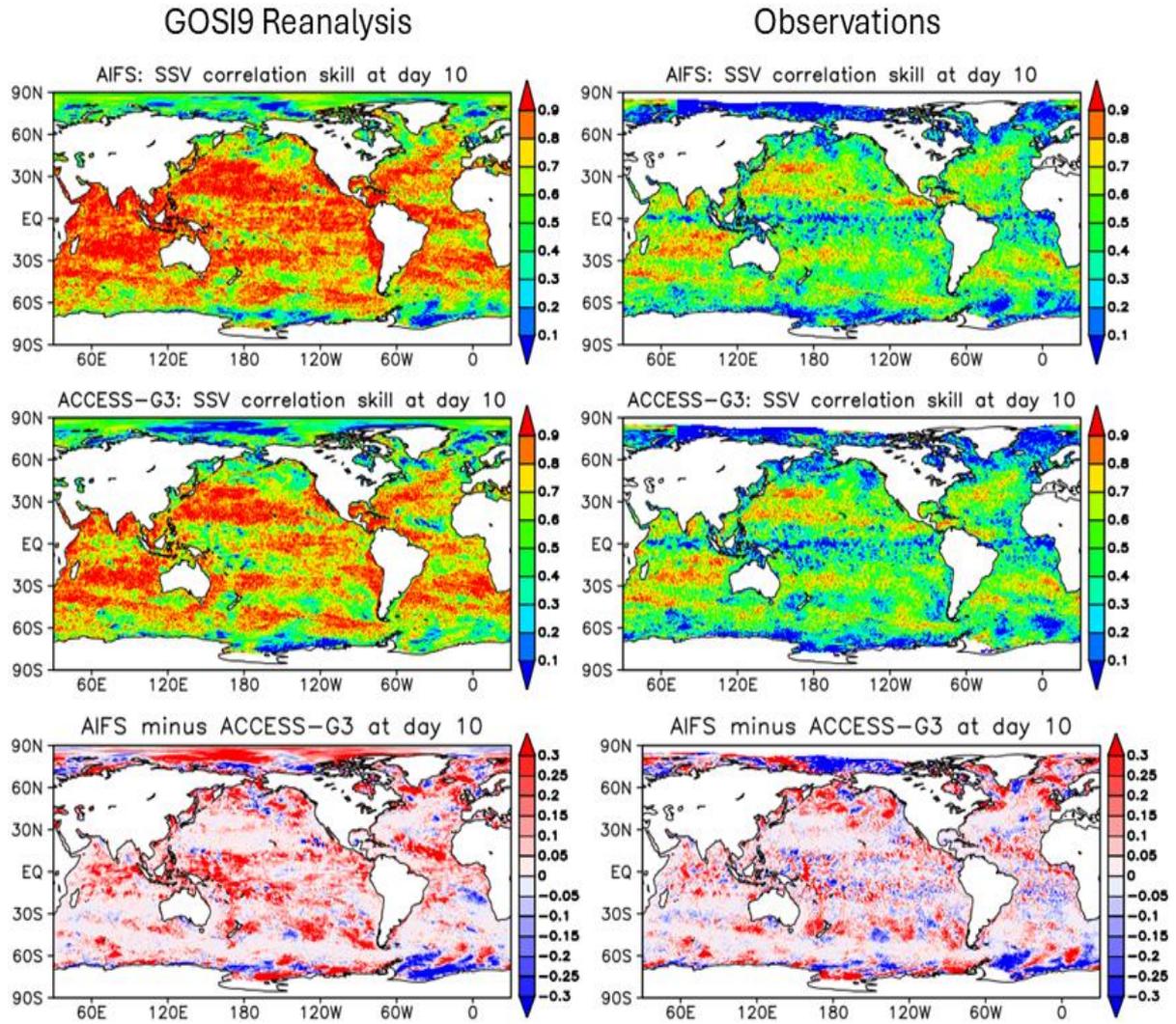

**Fig. 7.** Same as Fig. 6 except for sea surface meridional current (SSV).

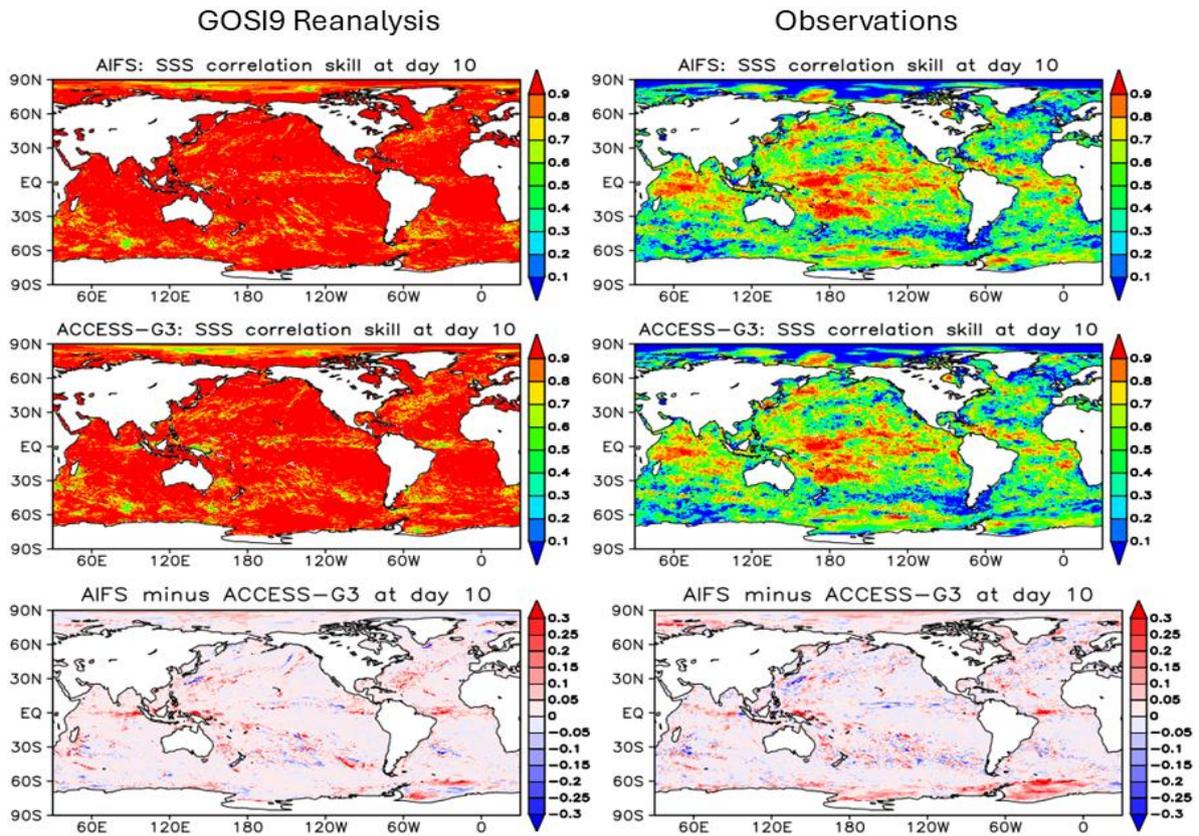

**Fig. 8.** Same as Fig. 6 except for sea surface salinity (SSS).

Figures 6–7 show that SSU and SSV forecasts exhibit high correlation skill in tropical and subtropical regions when evaluated against the GOSI9 reanalysis for both model experiments. In contrast, correlation skill declines in mid- and high-latitude regions. The elevated skill in the tropics reflects the strong and persistent influence of trade winds, which drive large, coherent, and predictable currents such as the North and South Equatorial Currents. This direct wind–current coupling enhances the predictability of surface currents in ocean models. In mid-latitudes, however, the variability of westerly winds and the influence of the Coriolis effect reduce forecast skill. Westerly winds are less stable and more variable than trade winds, and the Coriolis force deflects currents to the right in the Northern Hemisphere and to the left in the Southern Hemisphere. These deflections contribute to the formation of complex features such as meanders and mesoscale eddies, which are inherently chaotic and more challenging to predict.

When evaluated against observational data rather than the GOSI9 analysis, SSU and SSV again generally exhibit strong correlation skill across tropical regions in both model experiments. However, SSV shows notably low correlation near the equator. This discrepancy may be attributed to the dominance of zonal currents in equatorial dynamics, whereas meridional currents are comparatively weaker and more closely associated with tropical instability waves, which are inherently less predictable (Halpern et al., 1988). Additionally, the observational sea surface current data used for evaluation are only partially based on direct measurements. Due to the limited availability of surface current observations from drifters and coastal radars, the dataset integrates both satellite-derived geostrophic currents and modelled Ekman currents. The geostrophic component is derived from altimeter data, while the Ekman currents are generated using a model driven by ERA5 reanalysis winds. Consequently, model-related uncertainties may be introduced into the dataset (Rio et al. 2014).

At a 10-day lead time, sea surface current predictions (SSU and SSV) using AIFS data demonstrate higher overall correlation skill across most regions compared to those using ACCESS-G3 data. This improvement is evident in evaluations against both the GOSI9 reanalysis and independent observations. The enhanced correlation skill aligns with the superior performance of surface wind forecasts in AIFS (Figs. 1-4). Sea surface currents are strongly and directly influenced by surface winds; therefore, more accurate wind predictions generally lead to improved current forecasts.

For SSS, when the prediction skill is evaluated with respect to the GOSI9 reanalysis, the correlation is notably high, and substantially higher than when validated against independent observations (Fig. 8). It is important to note that satellite-derived SSS observations were not assimilated into the ocean model for initialization due to higher errors in satellite derived salinity observations (Brassington and Divakaran 2008). The spatial pattern of correlation between the model's reanalysis (not shown) and the observations during 2023–2024 closely resembles that between the forecasted SSS and the observations, highlighting the critical role of initial conditions in SSS forecasts. In general, SSS exhibits high persistence in most open ocean regions, particularly those far from freshwater sources such as river outflows and ice melt. This persistence arises because SSS is primarily governed by large-scale evaporation–precipitation balances and slow ocean circulation processes, which evolve over longer timescales. Across most ocean basins, SSS correlations are slightly higher in forecasts driven by AIFS compared to those forced with ACCESS-G3 data, regardless of whether the evaluation is conducted using GOSI9 reanalysis or independent observations.

Figures 9–12 show the RMSE for model-predicted SST, SSH, SSU, SSV and SSS at a 10-day forecast lead time. For ocean predictions driven by AIFS and ACCESS-G3, the spatial RMSE patterns are broadly similar between the two experiments for a given variable and lead time (Figs. 9-12; Figs. S3–S10).

For SST (Fig. 9), large forecast errors are concentrated in dynamically active regions including western boundary currents (Gulf Stream, Kuroshio, and Agulhas), and the Antarctic Circumpolar Current (ACC) in the Southern Ocean. In contrast, subtropical gyres and mid-latitude interior regions, where SST variability is governed by slower, large-scale processes, exhibit relatively low RMSE values, indicating higher predictability. The RMSE pattern for sea surface height anomaly (SSHA) closely resembles that of SST but displays more pronounced spatial structures (Fig. 9). Ocean forecasts driven by AIFS exhibit lower RMSE values for both SST and SSHA in the western tropical Pacific, tropical Indian Ocean, and eastern equatorial Pacific, compared to those forced with ACCESS-G3 data.

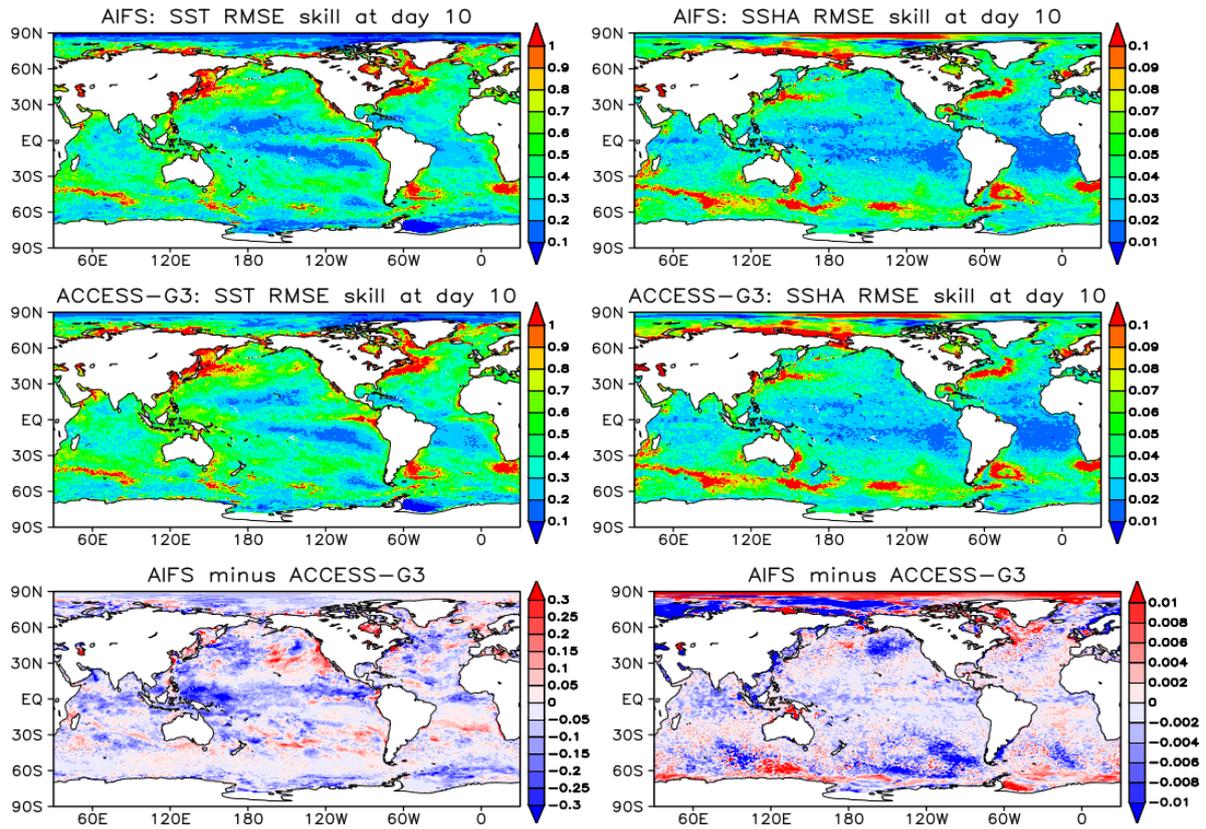

**Fig. 9.** Same as Fig. 5 except for RMSE. The units are ºC for SST and m for SSHA.

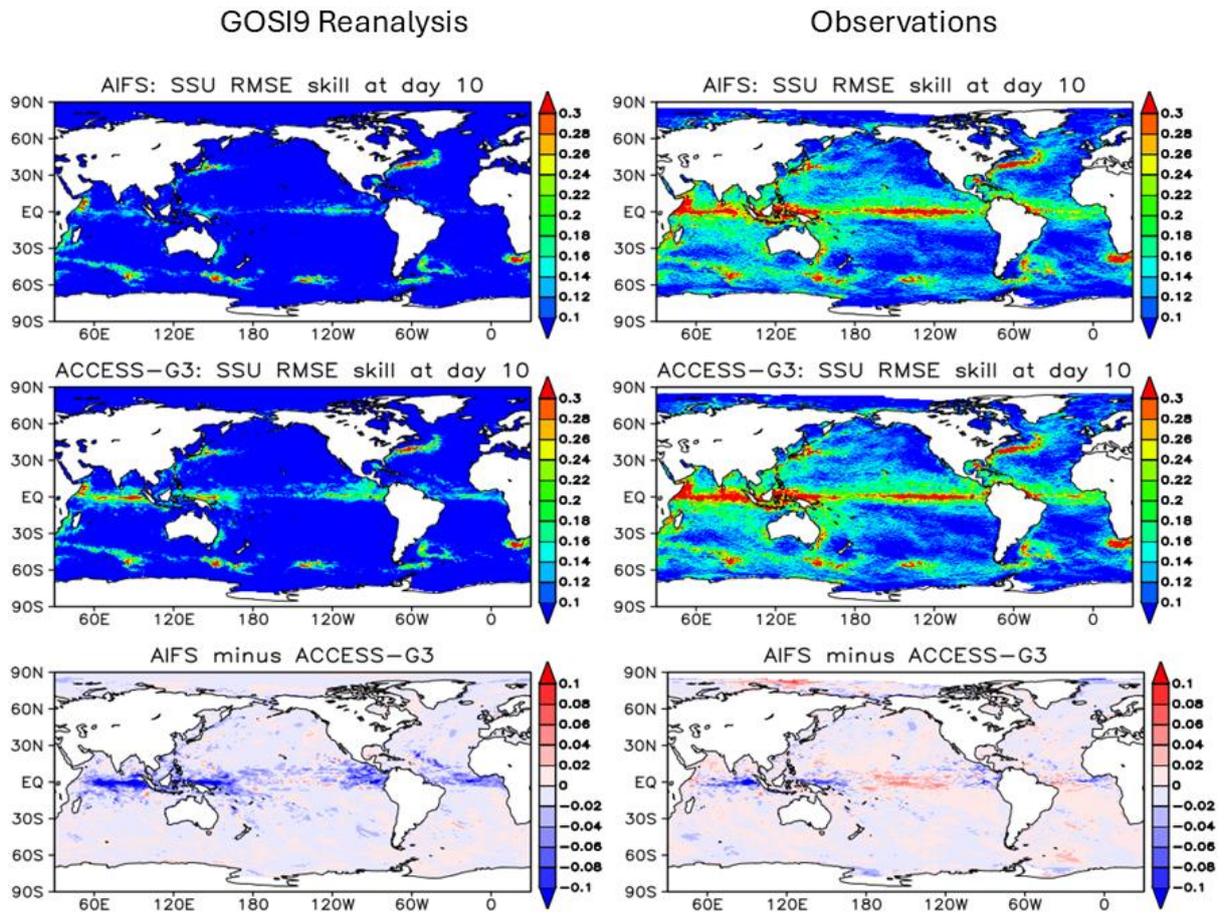

**Fig. 10.** Same as Fig. 6 except for SSU RMSE (unit: m/s).

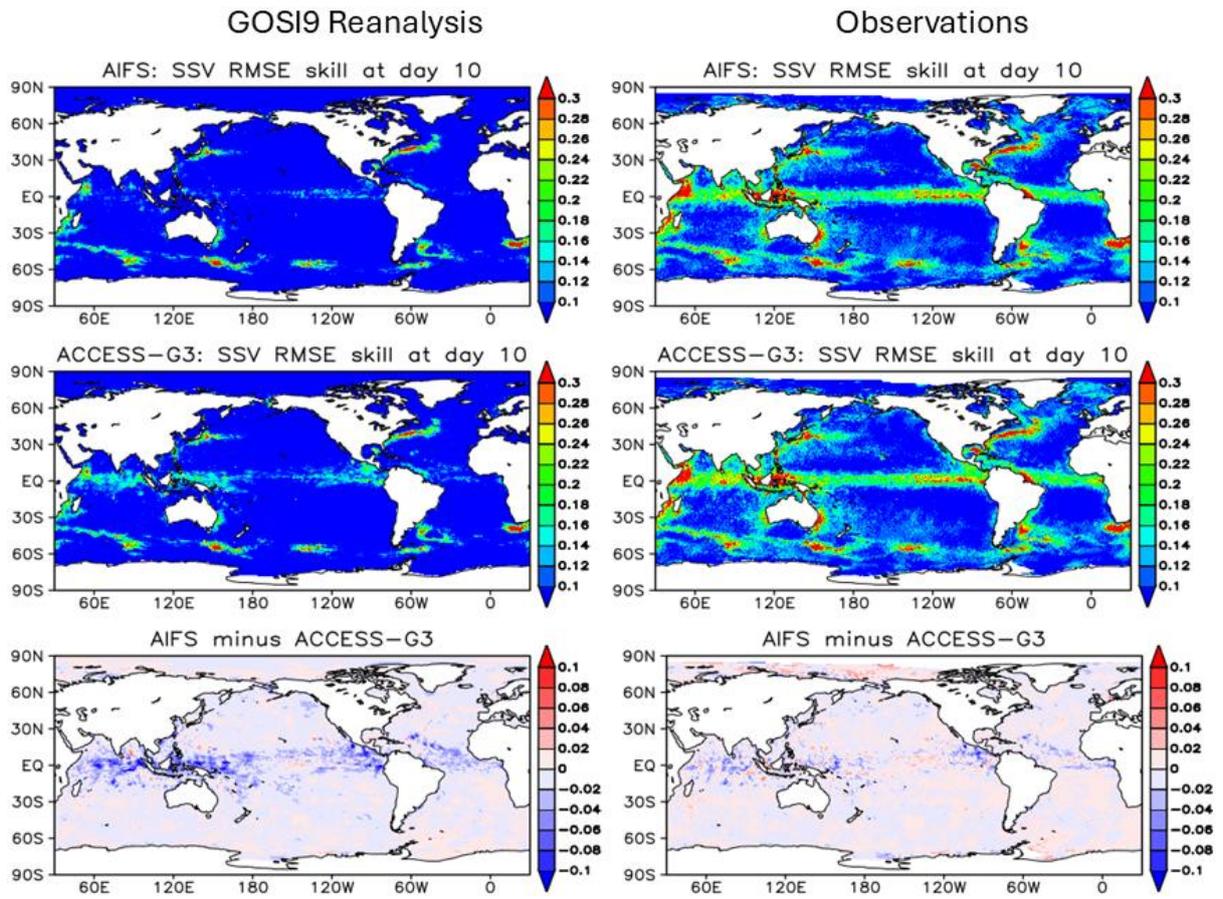

**Fig. 11.** Same as Fig. 7 except for SSV RMSE (unit: m/s).

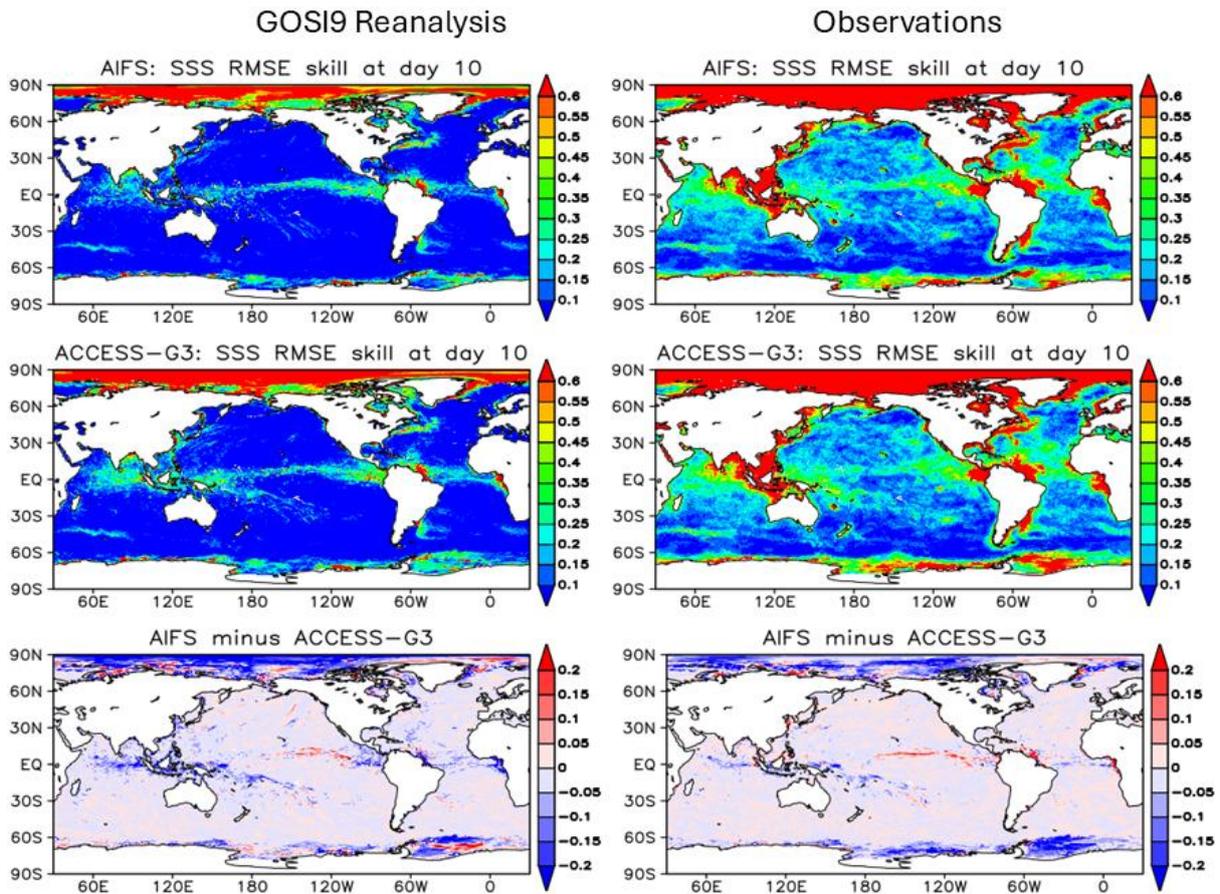

**Fig. 12.** Same as Fig. 8 except for SSS RMSE (unit: psu).

As expected, the RMSE values for SSU, SSV, and SSS (Figs. 10-12) are substantially lower when evaluated against the GOSI9 reanalysis than against independent observations, although the pattern of skill is similar. In AIFS-driven ocean forecasts, RMSE values for SSU and SSV are lower in tropical regions compared to those forced with ACCESS-G3 data, regardless of whether the evaluation is based on GOSI9 reanalysis or independent observations. These RMSE differences are more pronounced when assessed against GOSI9 reanalysis.

The RMSE distribution of SSS exhibits distinct spatial characteristics due to the dominant influence of freshwater fluxes. Elevated errors are observed in coastal and shelf regions affected by river discharge, as well as in equatorial zones—particularly within the ITCZ and SPCZ—where precipitation–evaporation imbalances strongly regulate surface salinity. High-latitude regions also display substantial RMSE, reflecting the impact of sea-ice interactions and uncertainties in freshwater forcing. Notably, even when SSS predictions are evaluated against the GOSI9 reanalysis, significant errors persist in the Arctic. In contrast, the open-ocean subtropical gyres generally exhibit low RMSE values, indicative of more stable salinity patterns and lower short-term variability. SSS forecasts forced with AIFS data exhibit lower RMSE values in areas near the Maritime Continent and the polar regions, particularly in the Arctic, compared to those driven by ACCESS-G3. However, slightly higher RMSE values are observed along the central and eastern equatorial Pacific in AIFS-driven forecasts relative to those using ACCESS-G3 forcings.

## 4.2 Lead-time dependent global mean correlation and RMSE

Figs. 13–16 present the global mean correlation and RMSE metrics from forecast day 1 through day 10 when forced with AIFS (red lines) and ACCESS-G3 (blue lines) atmospheric forcing. An additional experiment is shown, in which the ocean forecasts were forced with ERA5 atmospheric forcing (black lines). Forcing the ocean with ERA5 reanalysis data provides an upper-limit benchmark for ocean forecast performance with "observed" atmospheric forcing.

The global mean correlation for SST remains consistently high across lead times, exceeding 0.9, for all three forecast experiments (Fig. 13). Although SST correlations are comparable between AIFS and ERA5, the RMSE values associated with AIFS forcing become notably larger than those from ERA5 after day 5. Forecasts driven by ACCESS-G3 forcing exhibit the highest RMSEs across all lead times.

Despite the assimilation of altimeter data during model initialization, the global mean correlations for SSH are consistently lower than those for SST, with values around 0.82 at forecast day 1 (Fig. 13). This is primarily due to weak agreement between model reanalysis and satellite observations in dynamically complex regions, such as near 30°S in the Southern Ocean and 50°N in the Northern Hemisphere, where mesoscale eddy activity dominates (He et al., 2023). Consistent with SST results, the experiment with ACCESS-G3 forcing yields the lowest SSH correlations and the highest RMSE values throughout the forecast period. Moreover, its predictive skill deteriorates more rapidly than the other experiments.

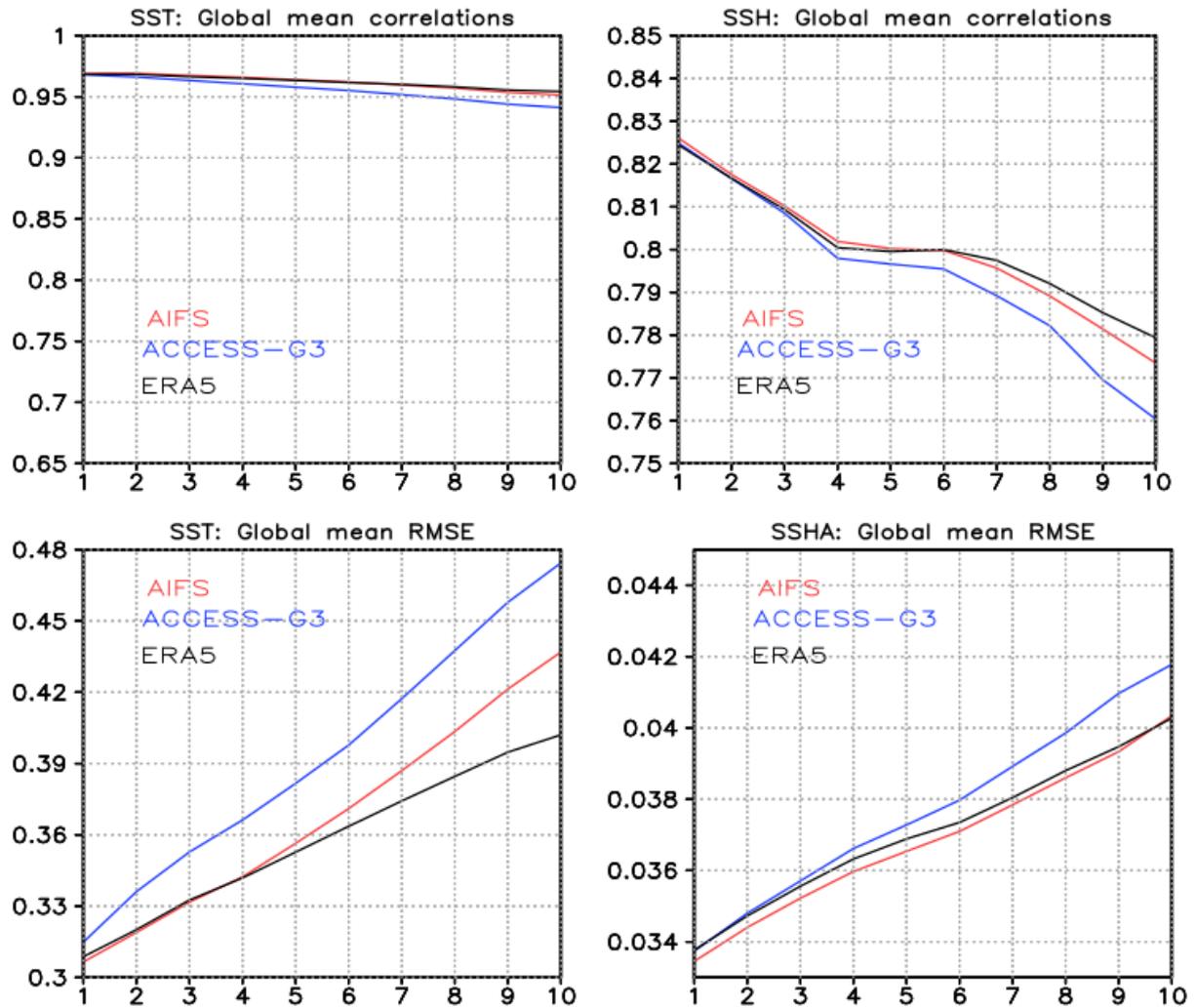

**Fig. 13.** Global mean correlations (top row) and RMSE (bottom row) from day 1 to 10 for SST (unit: °C; left column) and SSH (unit: m; right column). Ocean model forecasts forced with AIFS data are shown in red, and with ACCESS-G3 in blue. The black line shows ocean model results when forced with ERA5 data.

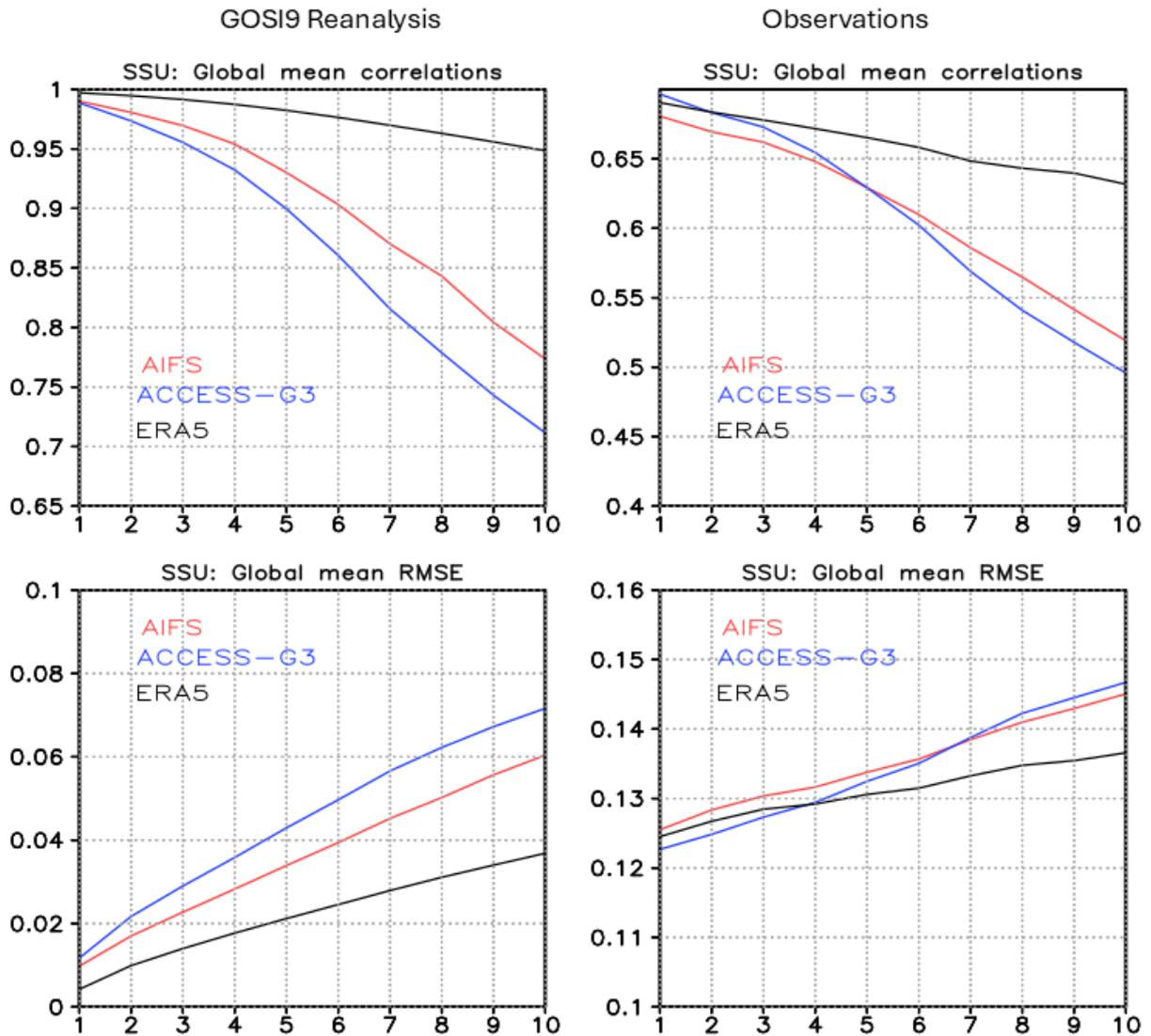

**Fig. 14.** Global mean correlations (top row) and RMSE (bottom row) from day 1 to day 10 for sea surface zonal current (SSU) (unit: m/s). The left column shows results evaluated against GOSI9 reanalysis data and the right column shows results evaluated against observations. Note the different y-axis scales between the left and right columns. Ocean model forecasts forced with AIFS data are shown in red, and with ACCESS-G3 in blue. The black line shows ocean model results when forced with ERA5 data.

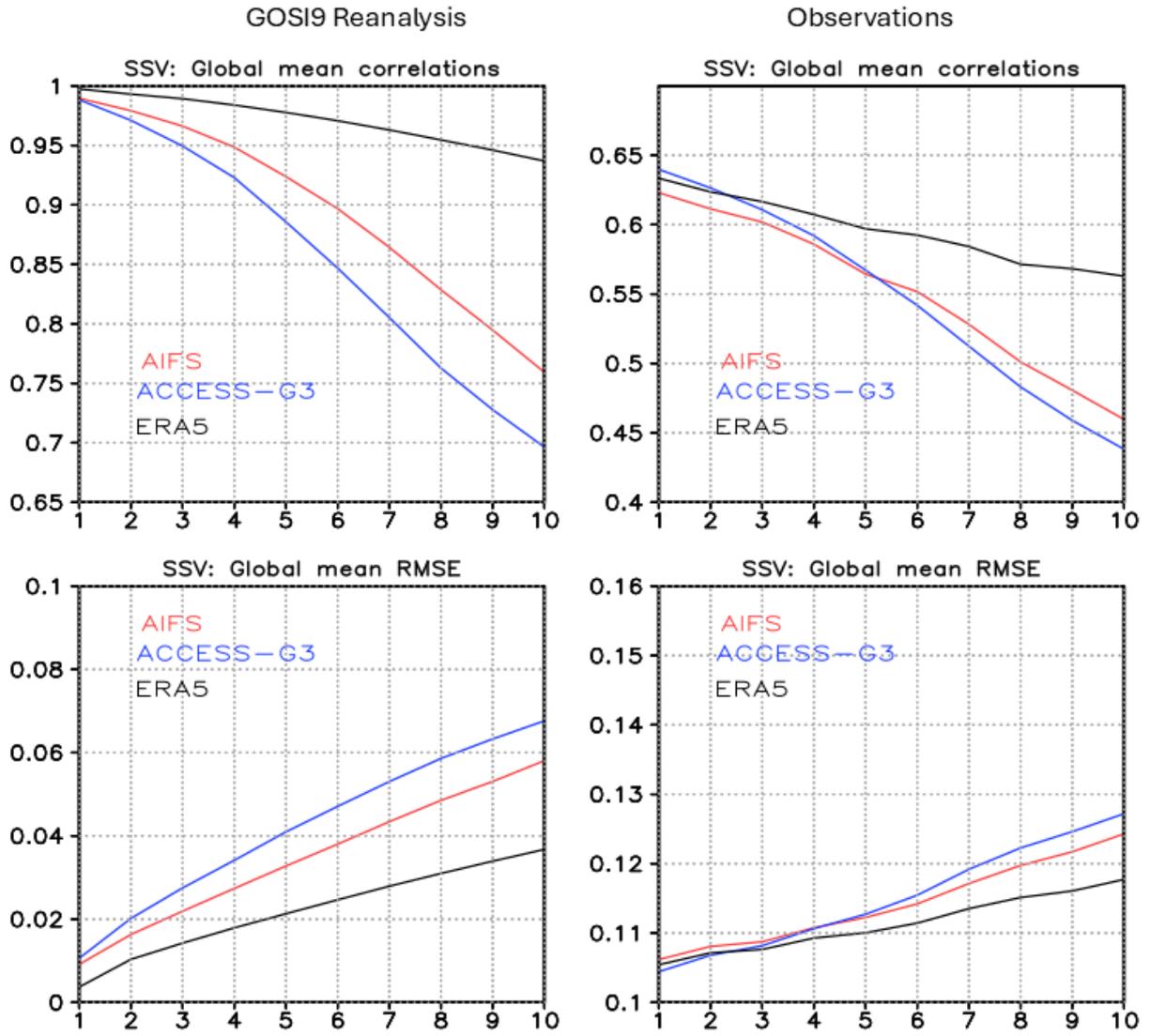

**Fig. 15.** Same as Fig. 14 except for sea surface meridional currents (SSV) (unit: m/s).

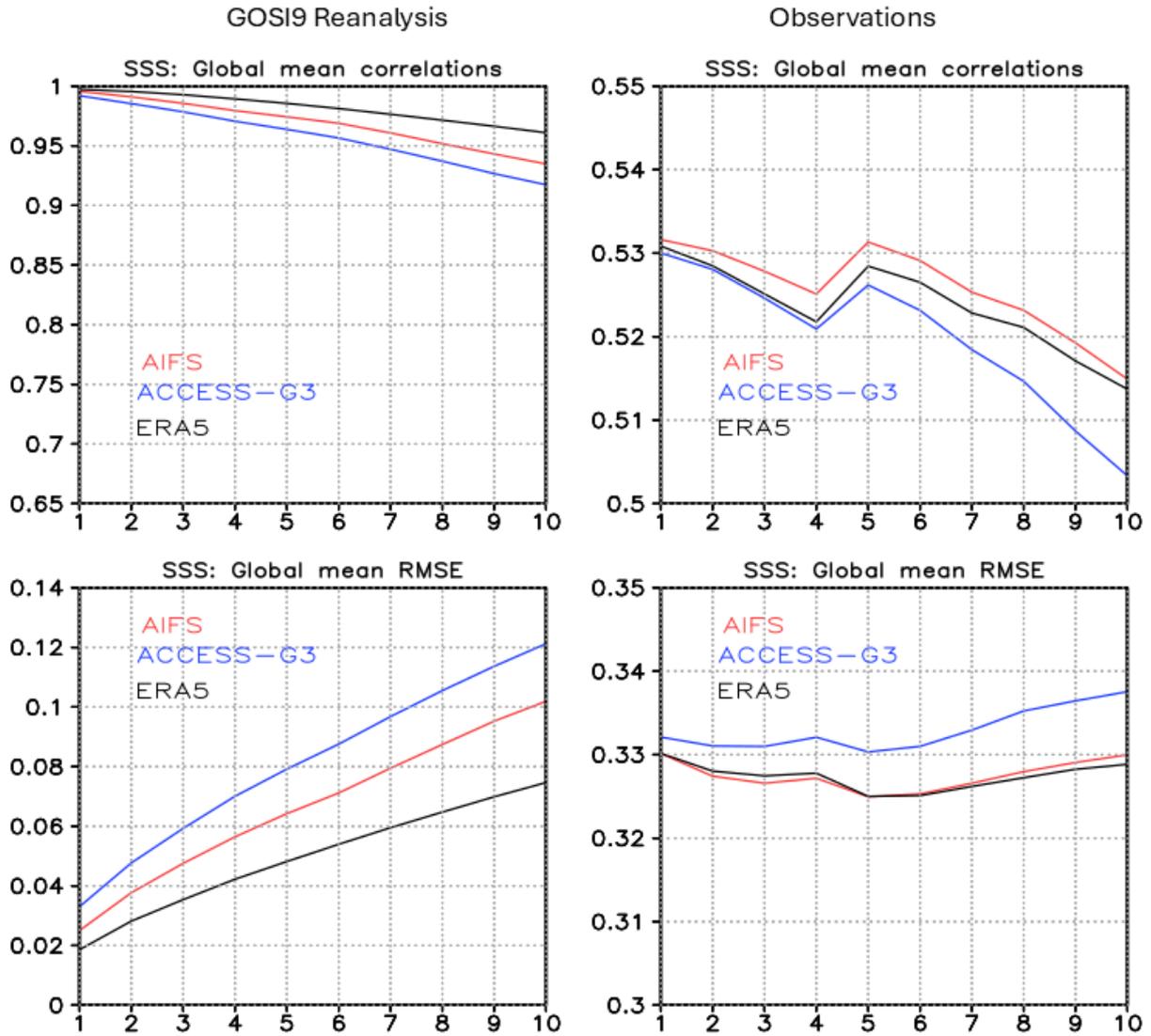

**Fig. 16.** Same as Fig. 14 except for sea surface salinity (SSS) (unit: psu).

Figures 14–16 present the global mean correlation and RMSE for SSU, SSV and SSS from forecast day 1 to day 10. At day 1, ocean forecasts driven by different atmospheric forcings exhibit similar performance, because they share the oceanic initial conditions from the GOSI9 reanalysis. When evaluated against the GOSI9 reanalysis, the global mean correlations at day 1 approach 1.0, indicating high agreement. In contrast, validation against independent observations yields significantly lower correlations at day 1—0.68 for SSU, 0.64 for SSV, and 0.53 for SSS.

When evaluated against the GOSI9 reanalysis, the AIFS-forced forecasts outperform those driven by ACCESS-G3 across all lead times. When evaluated against independent observations, AIFS-driven forecasts for SSU and SSV show slightly lower predictive skill than ACCESS-G3-driven forecasts during the first ~3 days, but subsequently outperforms at longer lead time. For SSS, AIFS-driven forecasts consistently outperform those based on ACCESS-G3 forcing.

## 5. Summary and discussion

This study has explored the viability of employing ML–based atmospheric forcing to drive physics-based ocean forecasts. We assessed the performance of the UK Met Office's Global Ocean and Sea Ice model (GOSI v9; Guiavarc'h et al., 2025) when forced with atmospheric forecasts from ECMWF's AIFS v1.0 ML weather model (Lang et al., 2024a; Moldovan et al., 2025) and the Bureau of Meteorology's ACCESS-G3 NWP physics-based model (Bureau of Meteorology National Operational Centre, 2019). By initializing the ocean model with identical ocean initial conditions, the influence of atmospheric forcing is isolated and the potential of using ML-based weather models for ocean prediction is evaluated. Additionally, forecasts driven by ERA5 reanalysis (Hersbach et al., 2020) served as an upper-limit benchmark representing performance under 'observed' atmospheric conditions.

Our evaluation of key atmospheric variables used to force the ocean model (e.g., 2-m temperature, mean sea level pressure, 10-m wind speed, downward shortwave radiation, and precipitation) indicates that AIFS forecasts generally outperform those from ACCESS-G3. This is evident from day 1 when compared against ERA5, and from ~3 days when evaluated against the ACCESS-G3 analysis. The findings are consistent with previous research, which indicates that data-driven ML weather models, such as AIFS (Lang et al., 2024a), Pangu-Weather (Bi et al. 2023) and GraphCast (Lam et al., 2023), typically achieve RMSE accuracy as good as or better than traditional physics-based medium-range weather prediction models for most variables. It is important to note that our assessment focuses on the larger scales and does not evaluate extremes or fine-scale spatial variability of the atmospheric forcing. Like most first-generation ML weather models, AIFS tends to smooth, or "blur" forecast fields as lead time increases (due to training using MSE as a loss function), leading to a loss of fine-scale detail at longer lead times (Lang et al. 2024a).

In terms of ocean forecasts, this study focuses on the sea surface variables, including SST, SSH, SSS and sea surface currents. SST and SSH forecasts were evaluated against observational datasets, while sea surface currents and SSS were assessed using both the GOSI9 reanalysis and independent observations, as these variables were not assimilated during model initialization. Results indicate that forecasts driven by AIFS consistently achieve higher correlation and lower RMSE compared to those forced by ACCESS-G3, reflecting AIFS's better skill in atmospheric prediction.

The atmospheric forcing fields for both AIFS and ACCESS-G3 are linearly interpolated onto the ocean model grid to provide the forcing. However, case study evaluations have indicated that, even at a common 0.25° resolution, AIFS outputs are noticeably smoother than those from ACCESS-G3 (not shown). This is likely due to their coarser horizontal native resolution (0.25° for AIFS vs 12km for ACCESS-G3) and the "blurring" artifact noted earlier. The increased spatial smoothness of the AIFS forcing may favour smoother ocean responses, leading to lower RMSE in sea surface variables. Consequently, part of the apparent improvement in ocean forecast skill under AIFS forcing may stem from this smoothing effect rather than from superior atmospheric accuracy alone. Similarly, the use of 6-hourly forcing fields from AIFS, compared to 1-hourly fields from ACCESS-G3, may reduce high-frequency variability in the atmospheric forcing, potentially favouring a smoother ocean response and lower RMSE. However, this may be mitigated by the fact that the AIFS forcing is 6-hourly instantaneous data, whilst the ACCESS-G3 forcing is hourly mean data. These factors will be examined in future work, including experiments with the next-generation AIFS-ENS ensemble

ML model (Lang et al. 2024b), which mitigates the "blurring" issue and enables ensemble-based investigations.

There are further limitations to this study that will be addressed in future work. Firstly, operational ocean forecasting systems typically employ higher-resolution ocean models (~1/12°), whereas this study uses a coarser 0.25° configuration. It remains uncertain whether ML-based atmospheric forcing at 0.25° resolution will be sufficient to drive skilful forecasts at operational scales. Secondly, the analysis was based on a relatively small sample size, which limits the robustness of the conclusions. Expanding the number of forecast cases and testing across different seasons and regions will be essential to strengthen confidence in these findings.

This study has provided a global assessment of sea surface prediction skill, but future work will extend these results to more impactful applications, including extreme event prediction. Particular attention will be given to the East Australian Current and its associated mesoscale eddies in the Tasman Sea; to understand how ML-based atmospheric forcing influences the dynamics of this major western boundary current.

Overall, this study demonstrates the potential of ML–based weather forecasts to drive physics-based ocean models. Ocean forecasting systems could benefit not only from the improved accuracy of ML atmospheric forcing but also from the reduced latency these models offer, as they can run significantly faster than traditional physics-based NWP systems. In the future it is likely that ML ocean models will eventually become competitive with physical ocean models. In the meantime, leveraging ML weather models as an alternative source of high-quality, low-latency forcing represents a promising pathway for advancing operational ocean prediction.

**Conflicts of interest**

The authors declare that they have no conflicts of interest.

**Acknowledgments**

The computation for this work was performed using the NCI National Facility at the Australian National University. We thank Dr. Prasanth Divakaran and Dr. Chen Li for their comments on a pre-submission draft.

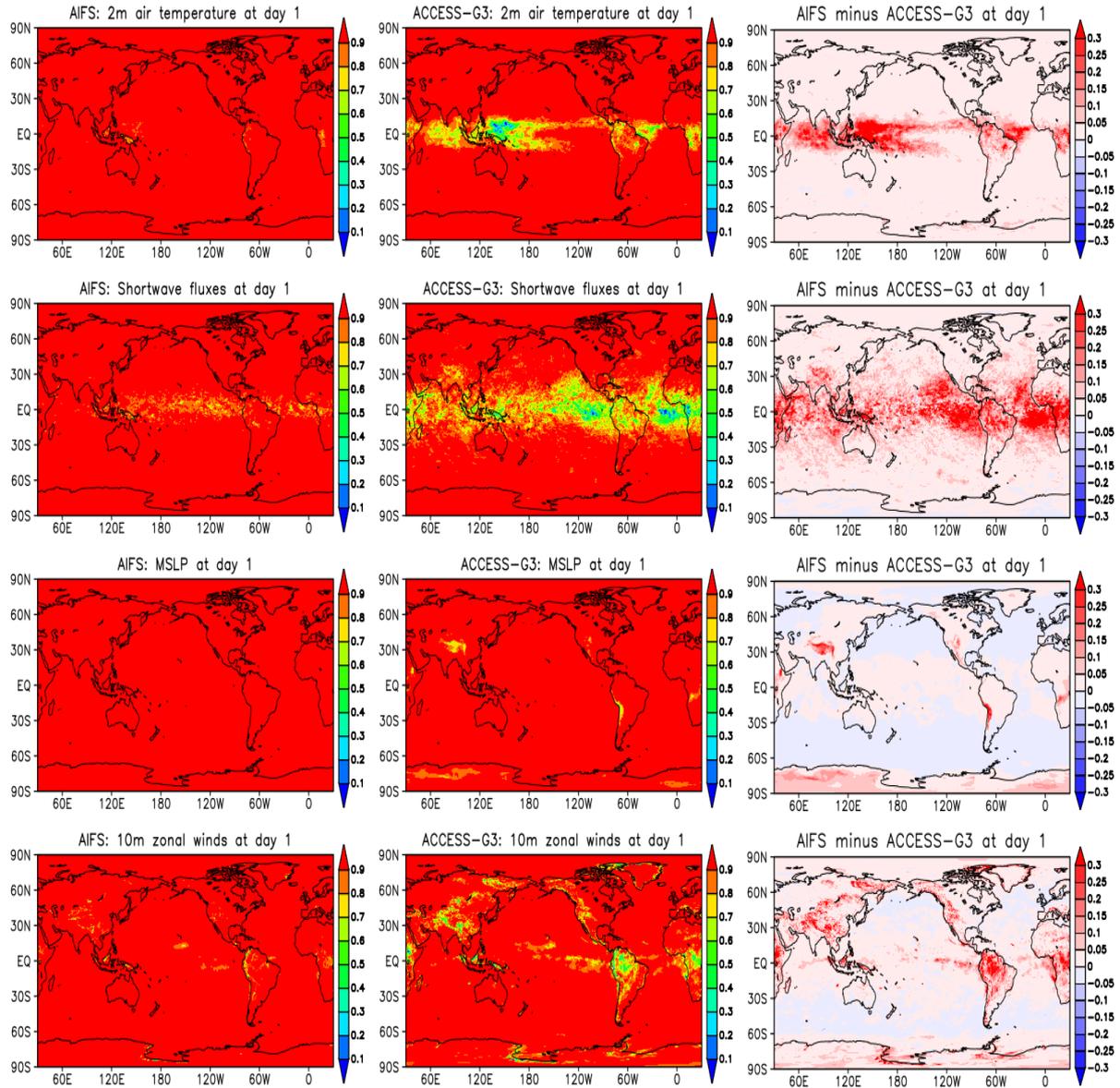

**Fig. S1.** Correlations between AIFS model forecasts and ERA5 reanalysis (left column), ACCESS-G3 forecast and ERA5 reanalysis (middle column) and the correlation differences between AIFS and ACCESS-G3 (right column) for day-1 of the forecasts. The atmospheric variables, from top to bottom, are 2m air temperature, surface downward shortwave flux, mean sea level pressure (MSLP) and 10m zonal wind.

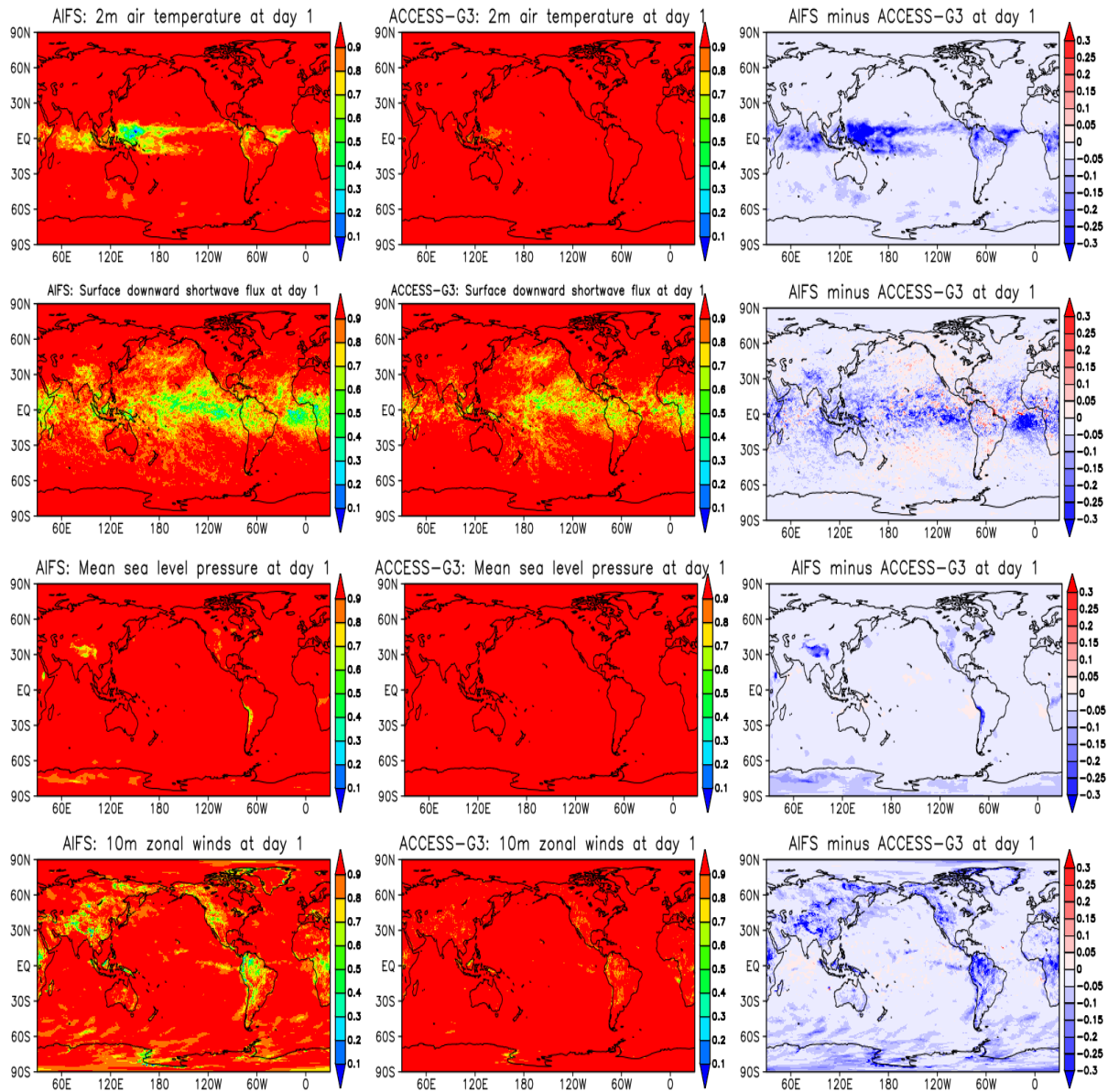

**Fig. S2.** Correlations between AIFS model forecasts and ACCESS-G3 analysis (left column), ACCESS-G3 forecast and ACCESS-G3 analysis (middle column) and the correlation differences between AIFS and ACCESS-G3 (right column) for day-1 of the forecasts. The atmospheric variables, from top to bottom, are 2m air temperature, surface downward shortwave flux, mean sea level pressure (MSLP) and 10m zonal wind.

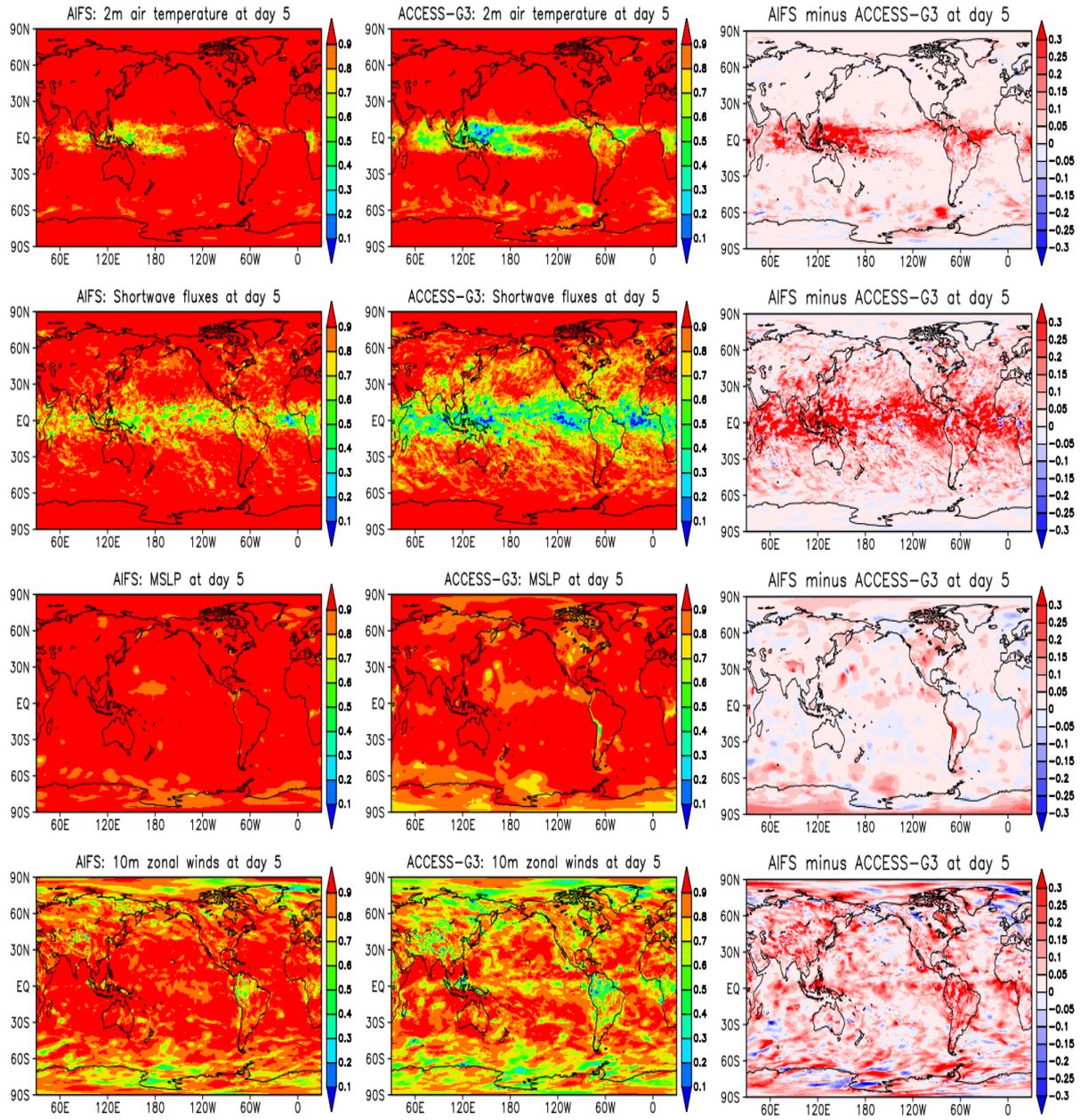

**Fig. S3.** Same as Fig. S1 but for day 5.

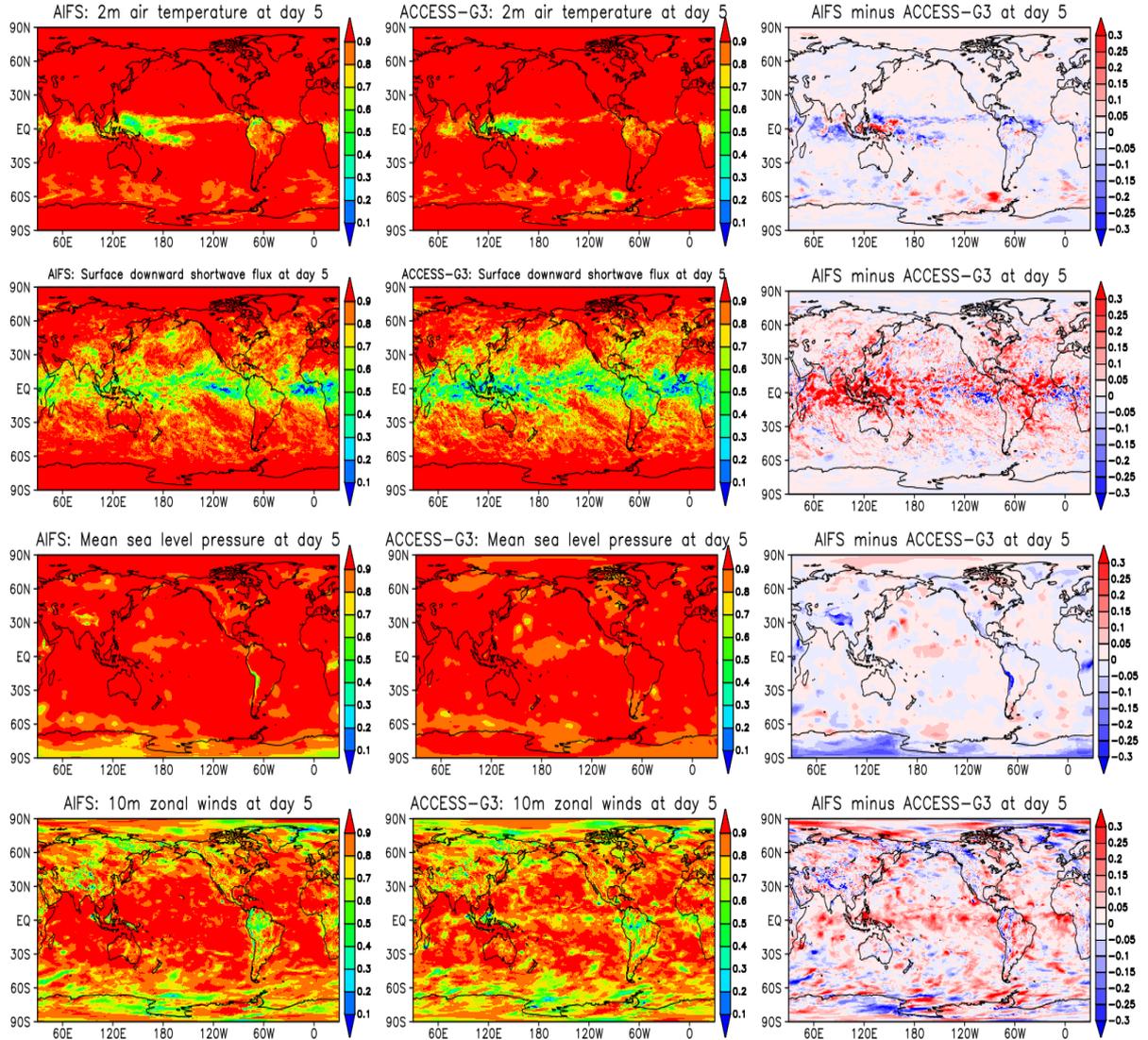

**Fig. S4.** Same as Fig. S2 but for day 5.

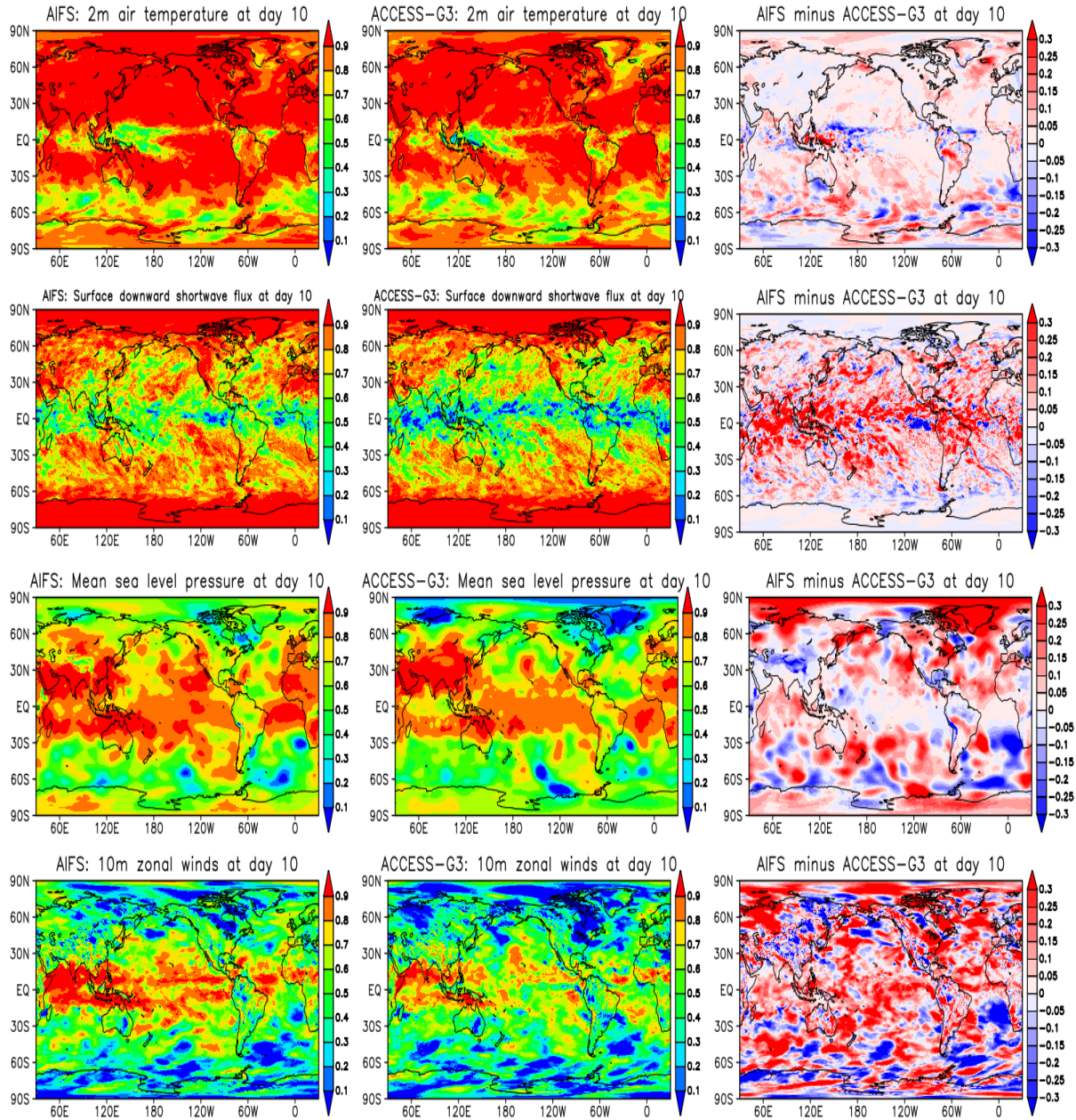

**Fig. S5.** Same as Fig. S2 but for day 10.

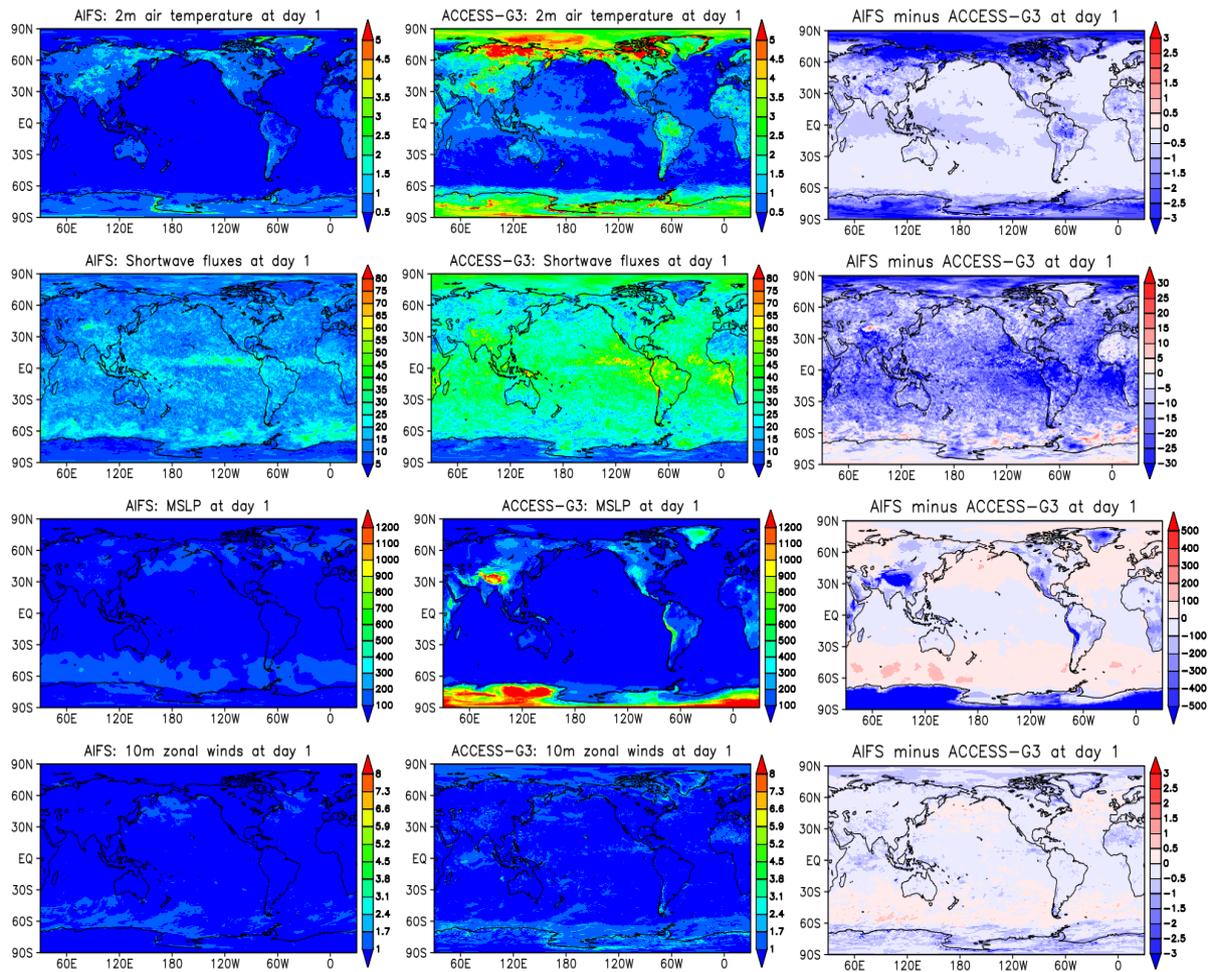

**Fig. S6.** RMSEs between AIFS model forecasts and ERA5 reanalysis (left column), ACCESS-G3 forecast and ERA5 reanalysis (middle column) and the RMSE differences between AIFS and ACCESS-G3 (right column) for day-1 of the forecasts. The atmospheric variables, from top to bottom, are 2m air temperature, surface downward shortwave flux, mean sea level pressure (MSLP) and 10m zonal wind. The units are K for air temperature, W/m$^2$ for short wave fluxes, Pa for MSLP and m/s for winds.

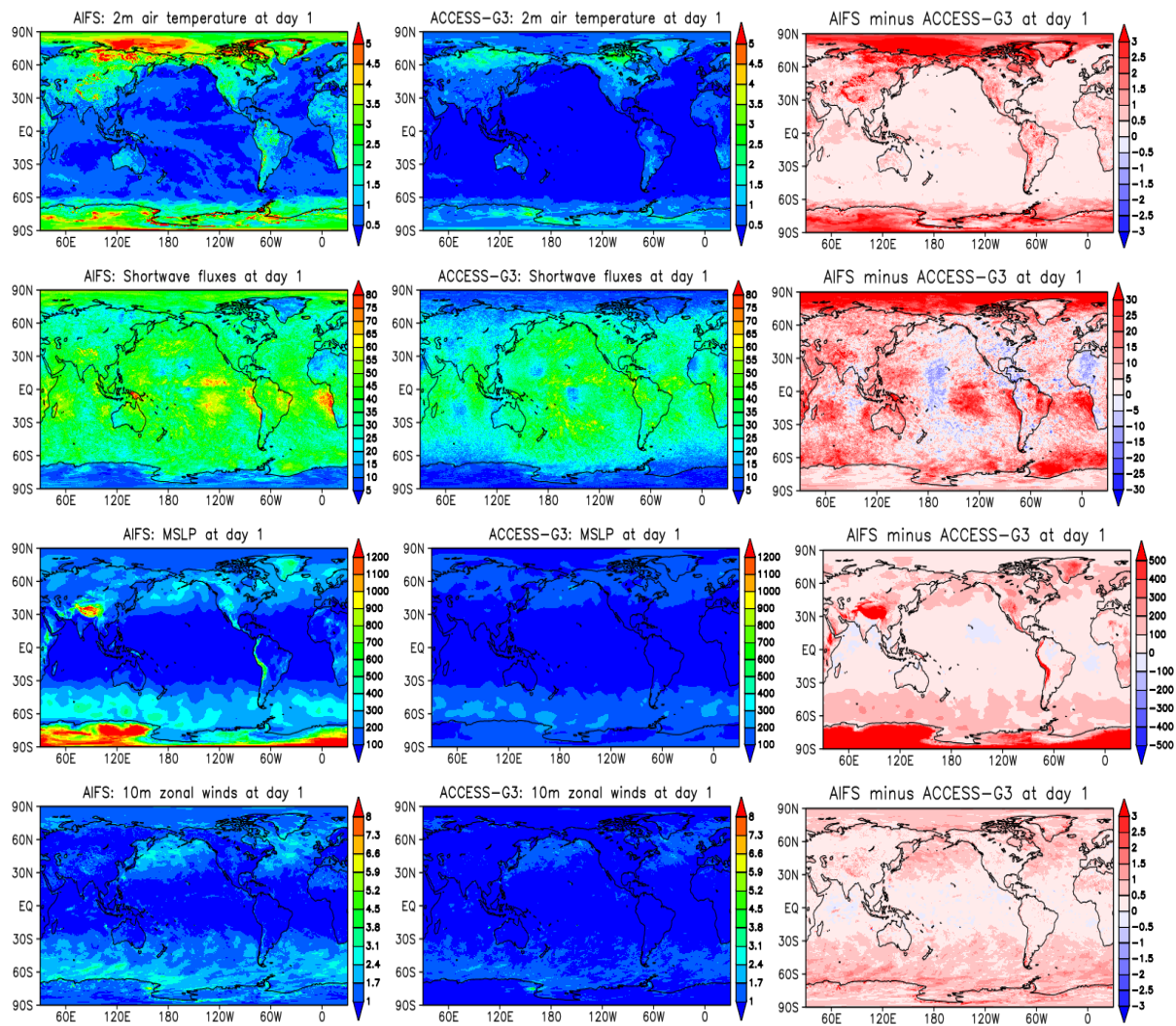

**Fig. S7.** RMSEs between AIFS model forecasts and ACCESS-G3 analysis (left column), ACCESS-G3 forecast and ACCESS-G3 analysis (middle column) and the RMSE differences between AIFS and ACCESS-G3 (right column) for day-1 of the forecasts. The atmospheric variables, from top to bottom, are 2m air temperature, surface downward shortwave flux, mean sea level pressure (MSLP) and 10m zonal wind. The units are K for air temperature, W/m² for short wave fluxes, Pa for MSLP and m/s for winds.

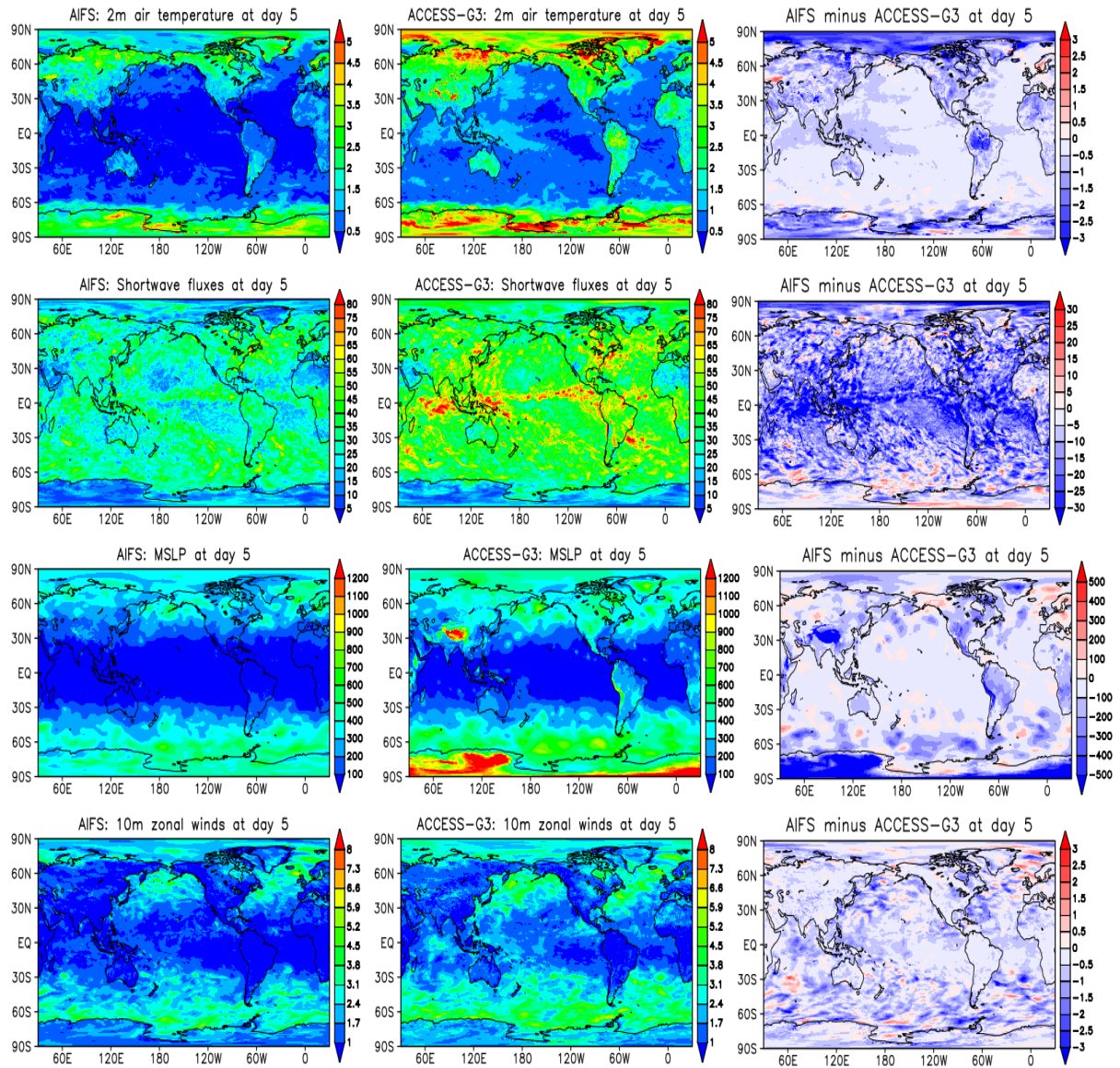

**Fig. S8.** Same as Fig.S6 but for day 5.

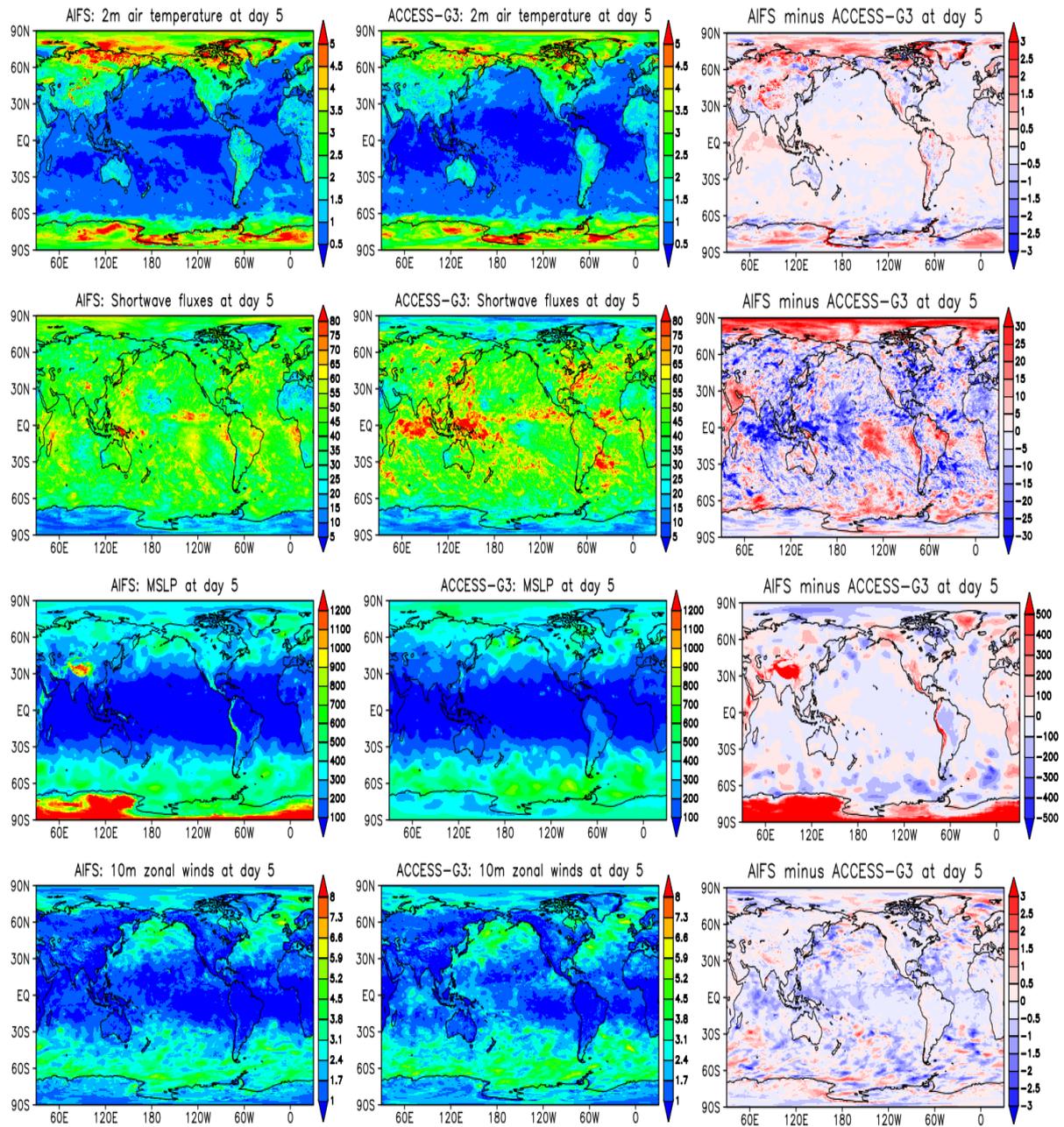

**Fig. S9.** Same as Fig. S7 but for day 5.

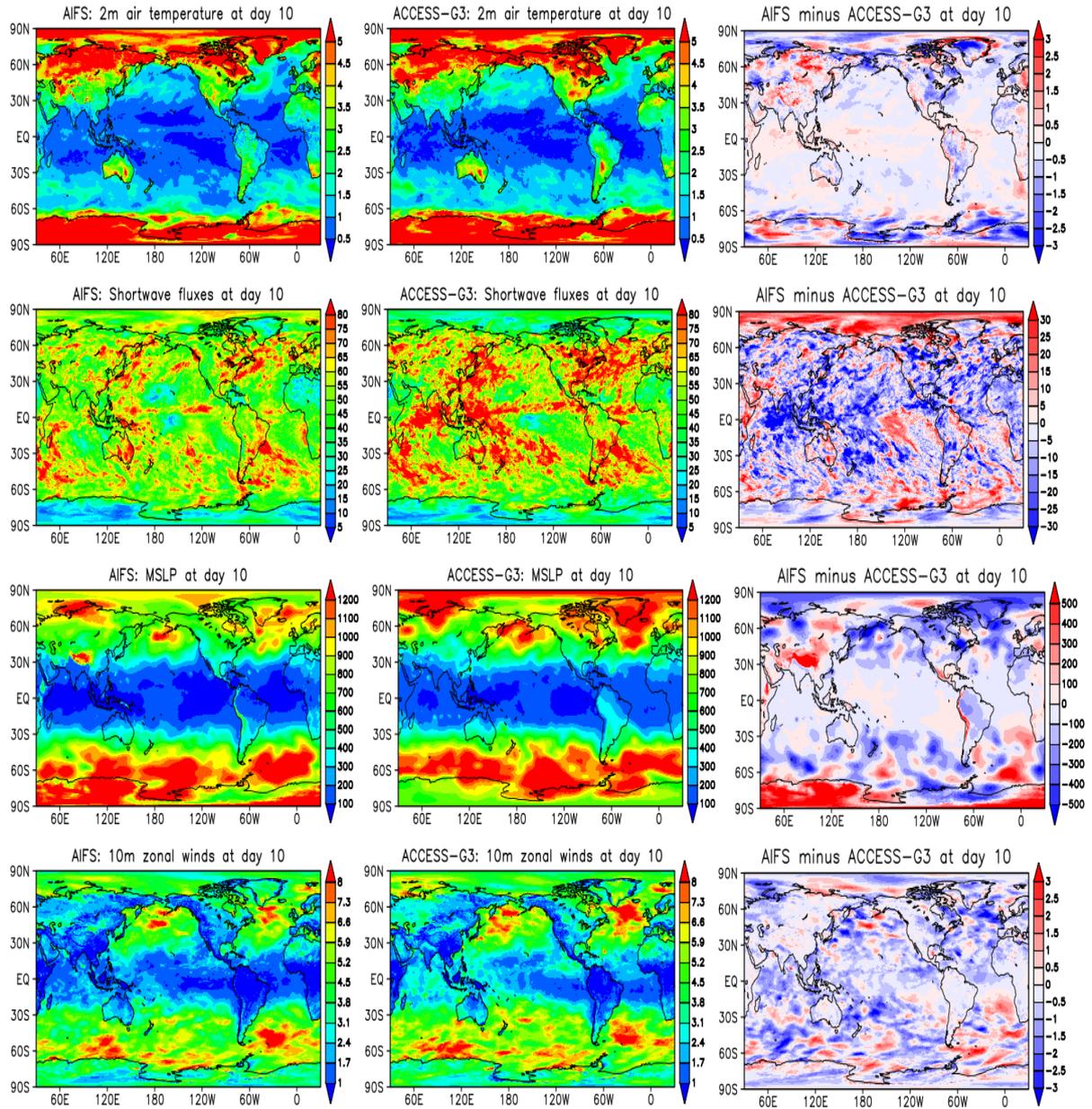

**Fig. S10.** Same as Fig. S7 but for day 10.

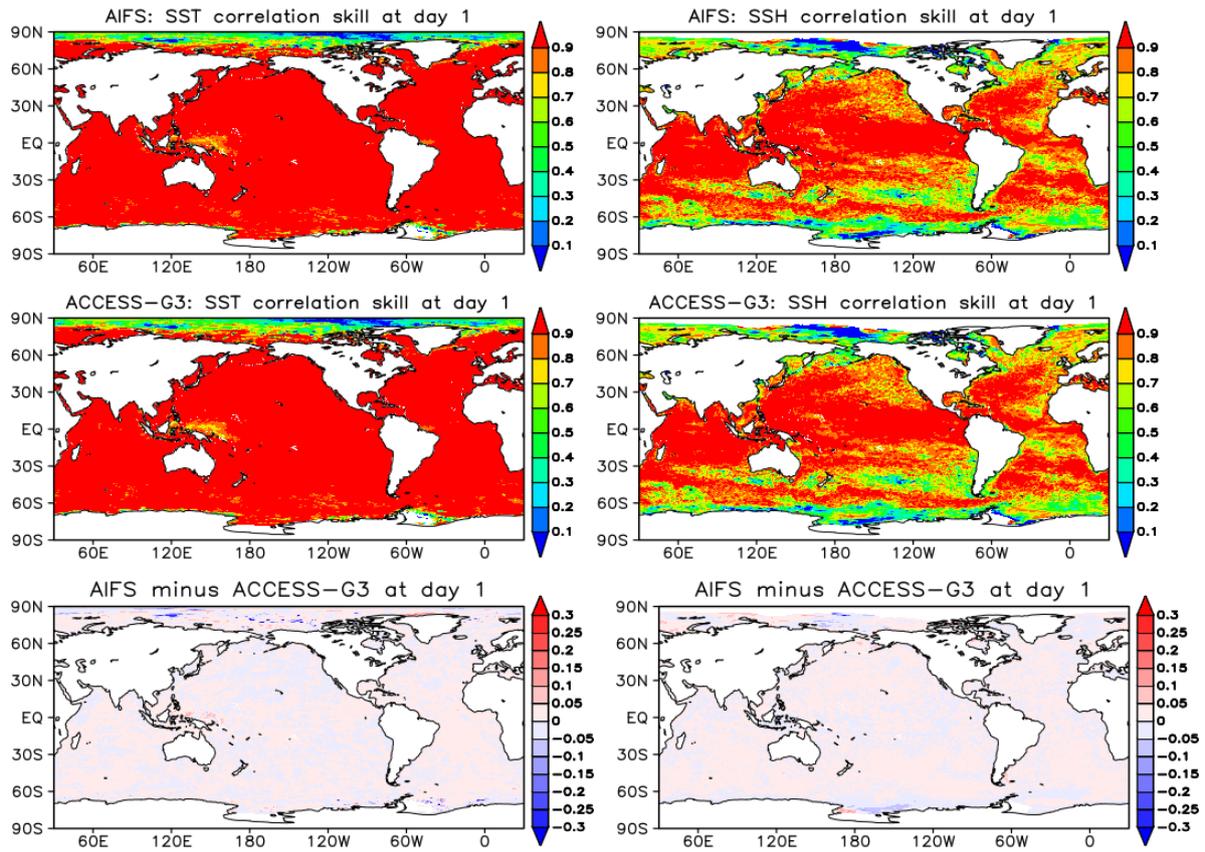

**Fig. S11.** Correlations between ocean forecasts and observations at day 1 for SST (left column) and SSH (right column). The top row shows forecasts forced with AIFS data, the middle row shows forecasts forced with ACCESS-G3 data and the bottom row shows the correlation differences between the two.

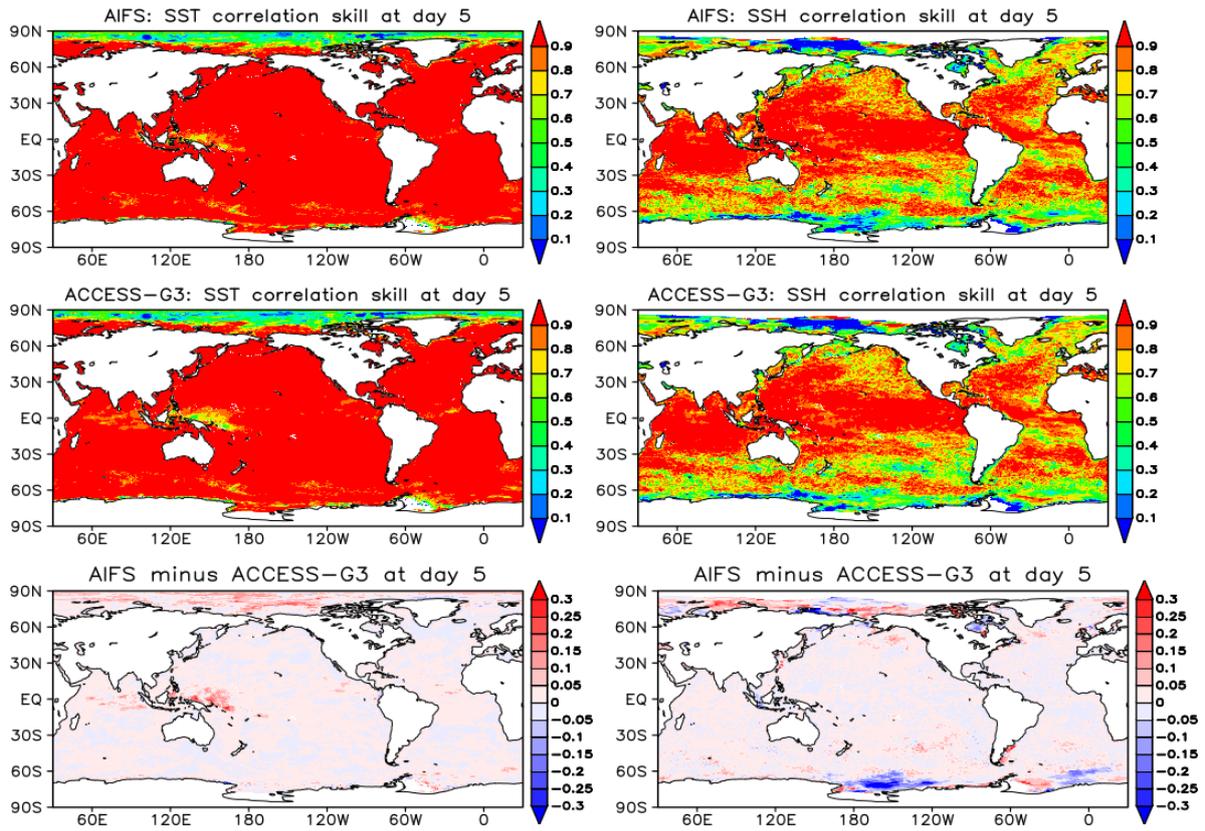

**Fig. S12.** Same as Fig. S11 but for day 5.

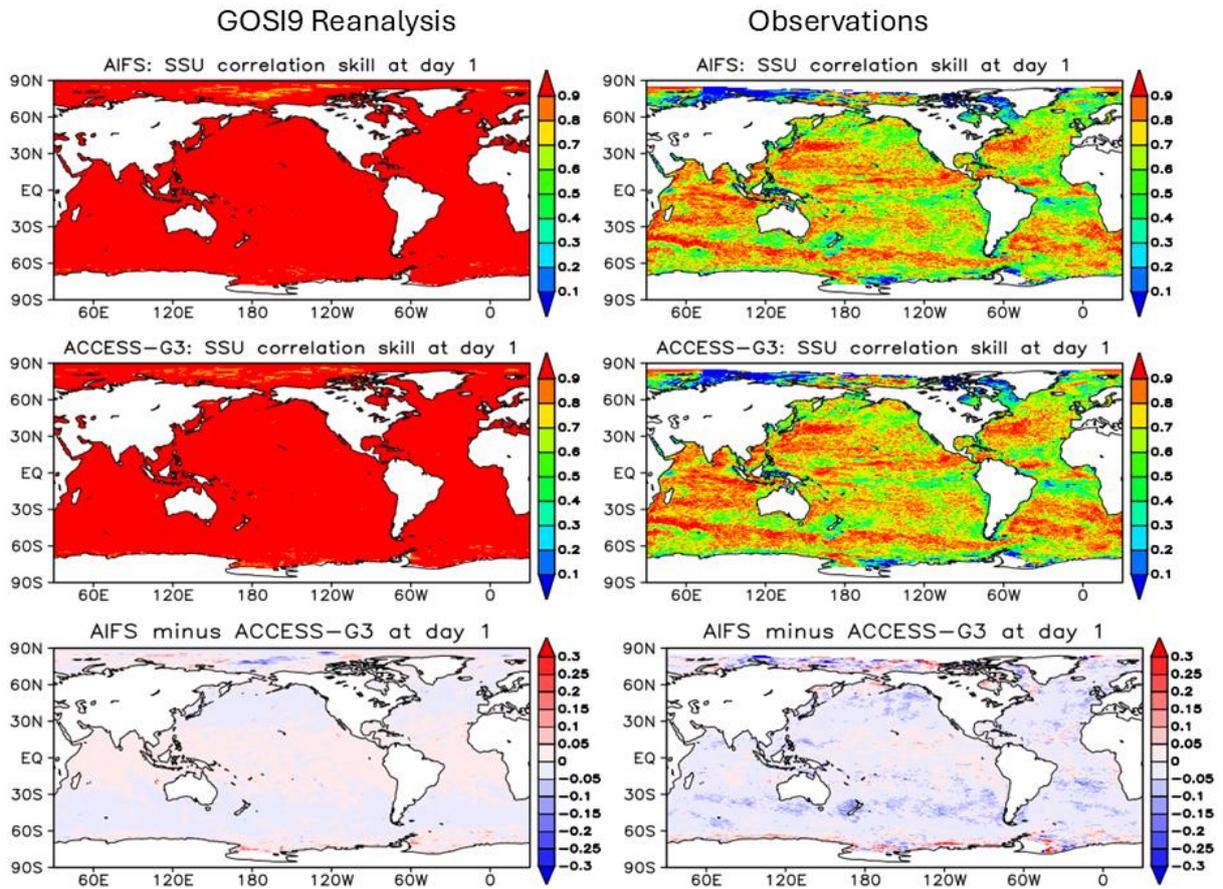

**Fig. S13.** Correlations at day 1 between model-forecasted sea surface zonal current (SSU) and GOSI9 ocean reanalysis (left column), and between model SSU and observations (right column). The top row shows forecasts forced with AIFS data, the middle row shows forecasts forced with ACCESS-G3 data and the bottom row shows the correlation differences between the two.

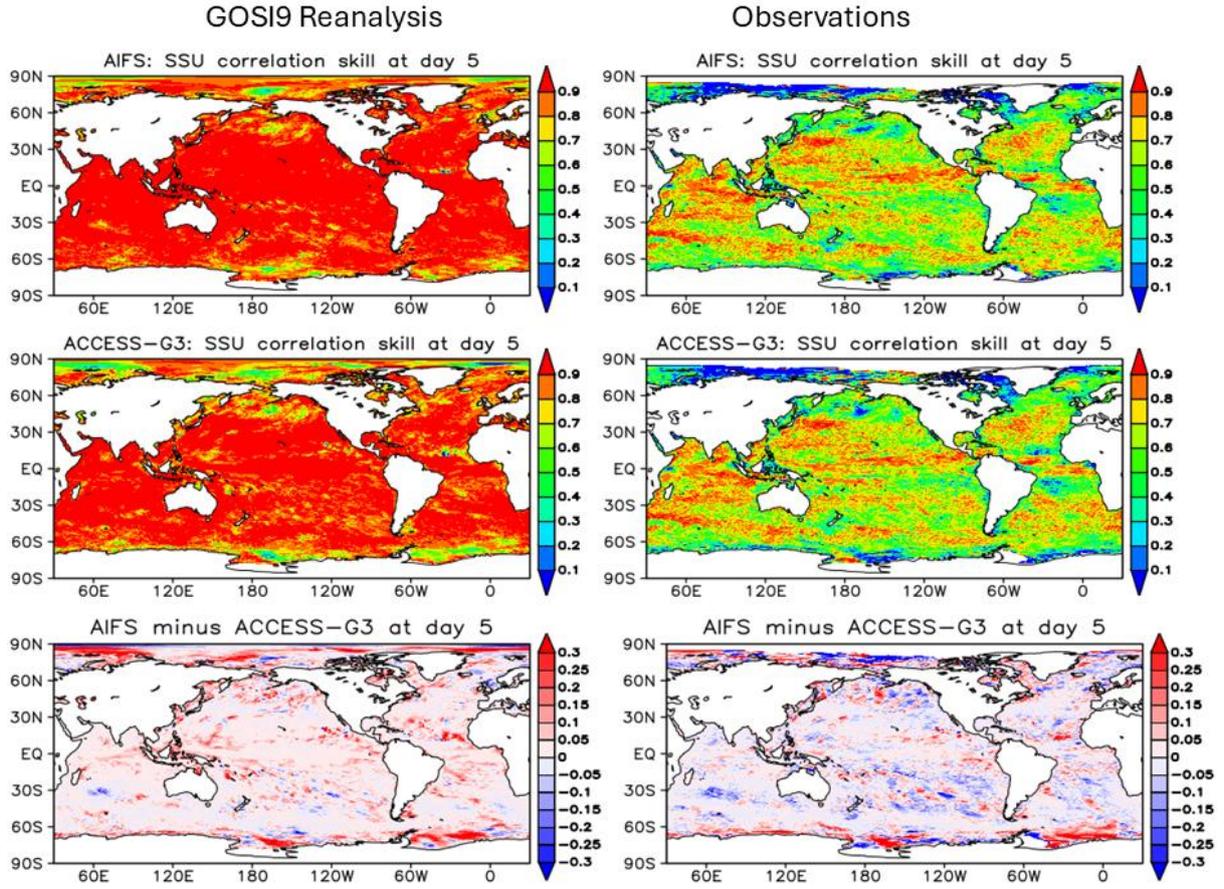

**Fig. S14.** Same as Fig. S13 but for day 5.

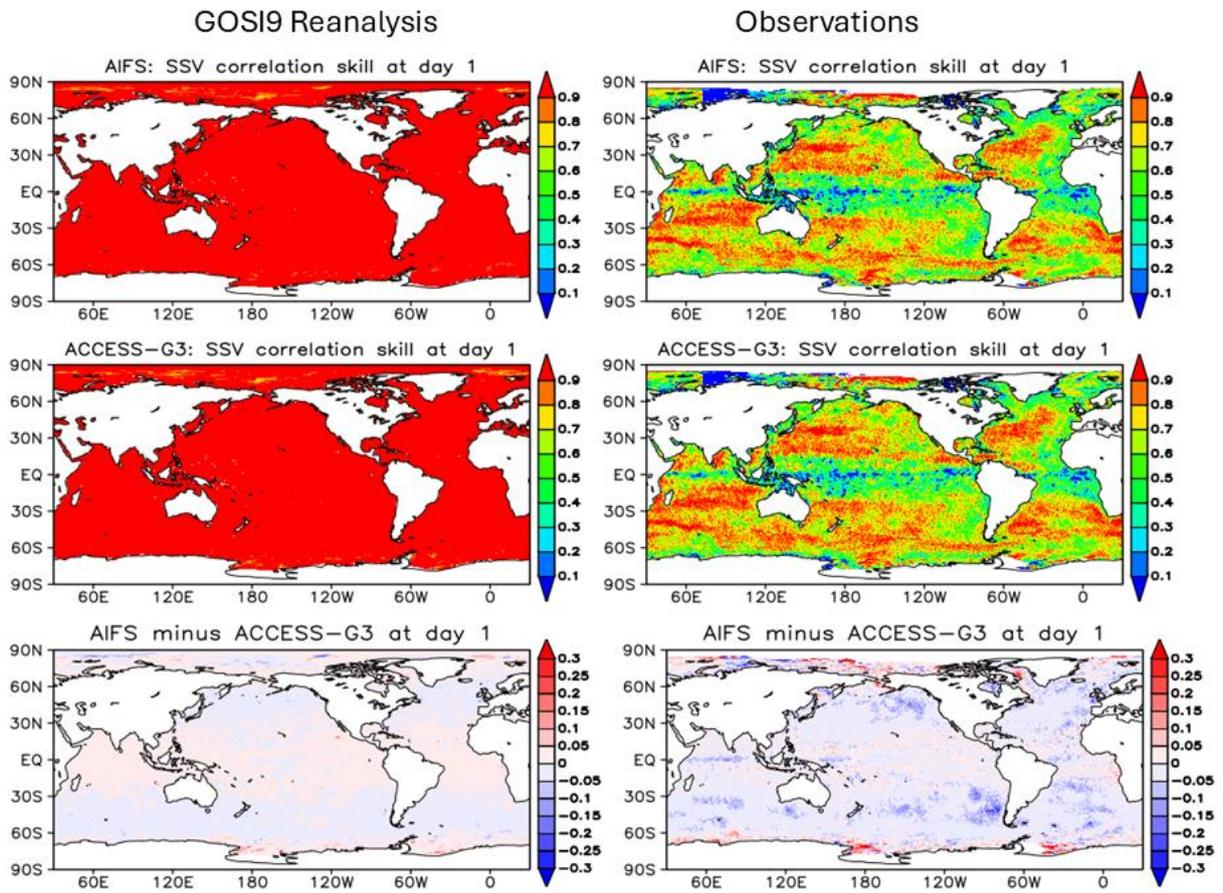

**Fig. S15.** Same as Fig. S13 but for sea surface meridional current (SSV).

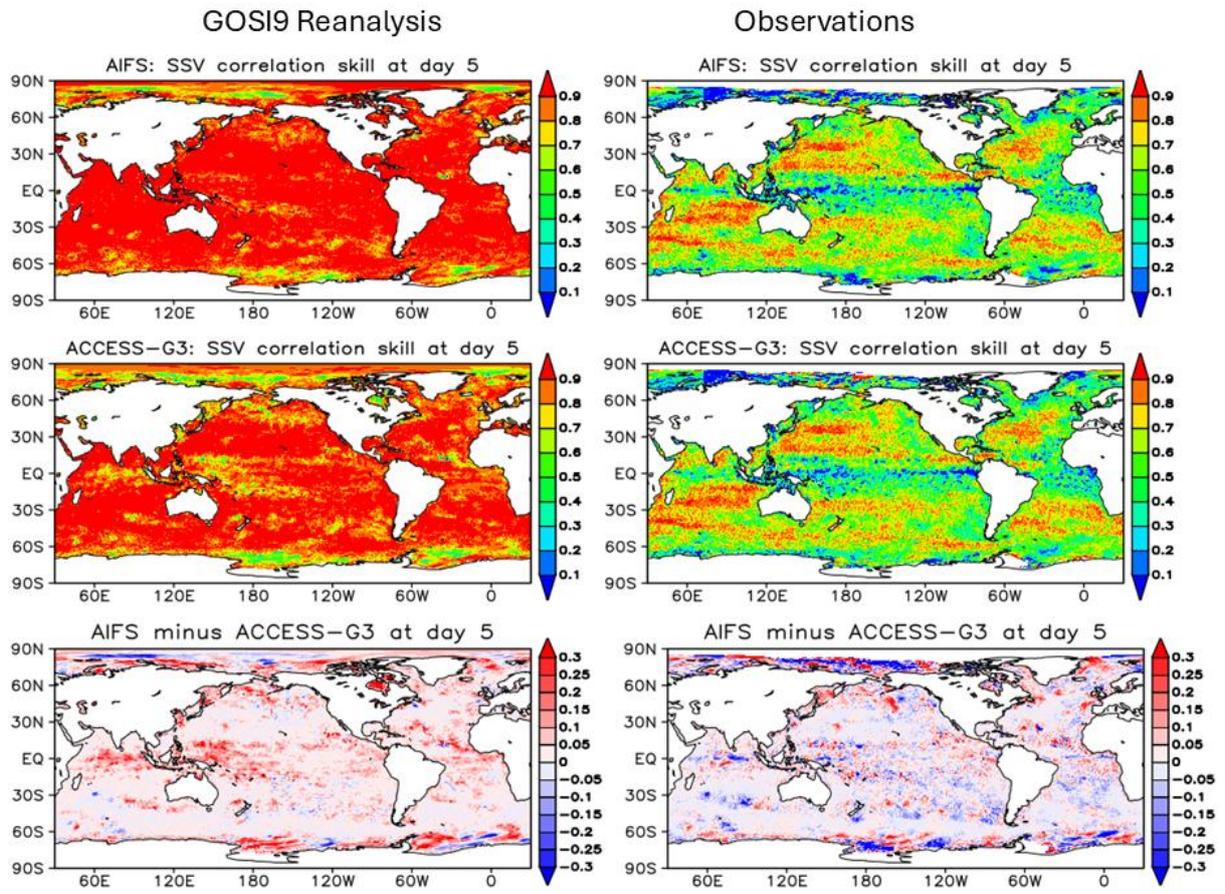

**Fig. S16.** Same as Fig. S14 but for day 5.

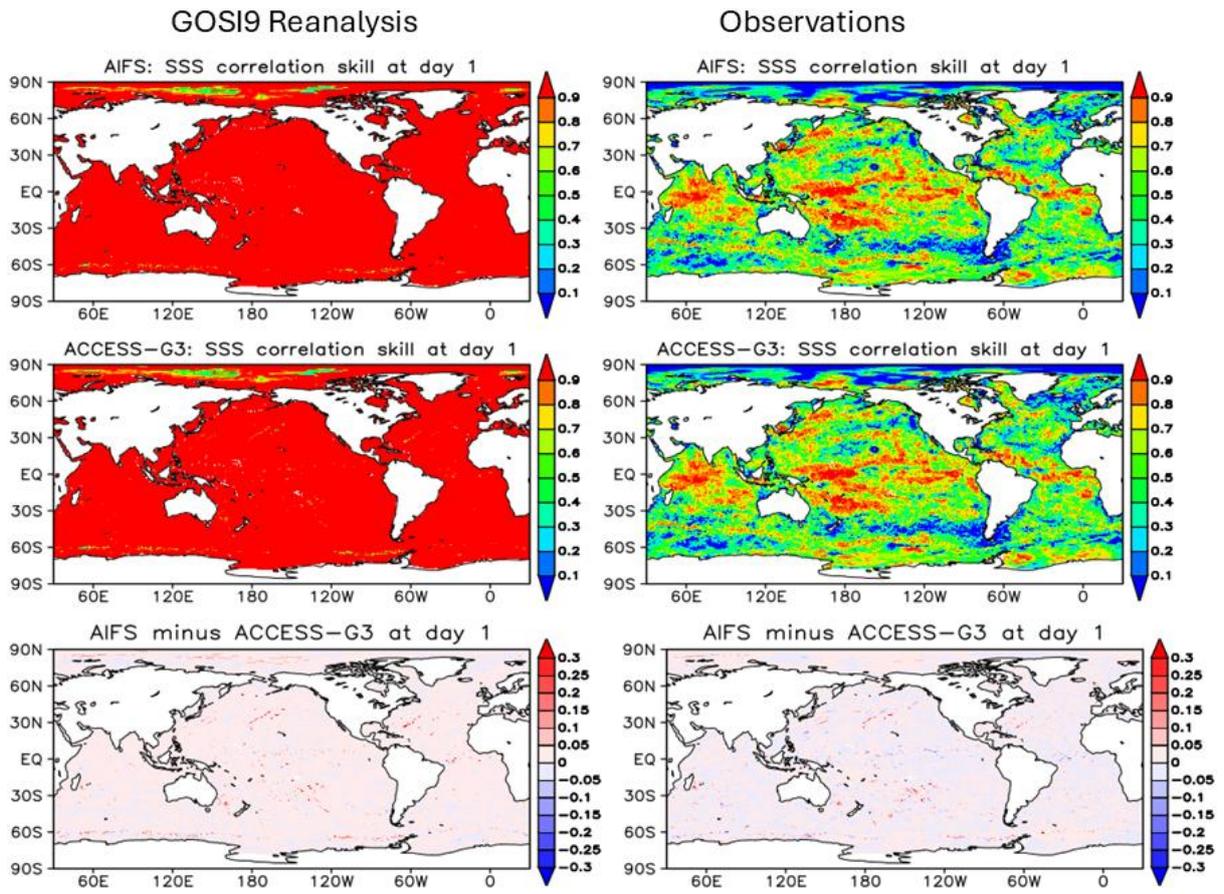

**Fig. S17.** Same as Fig. S13 but for sea surface salinity (SSS).

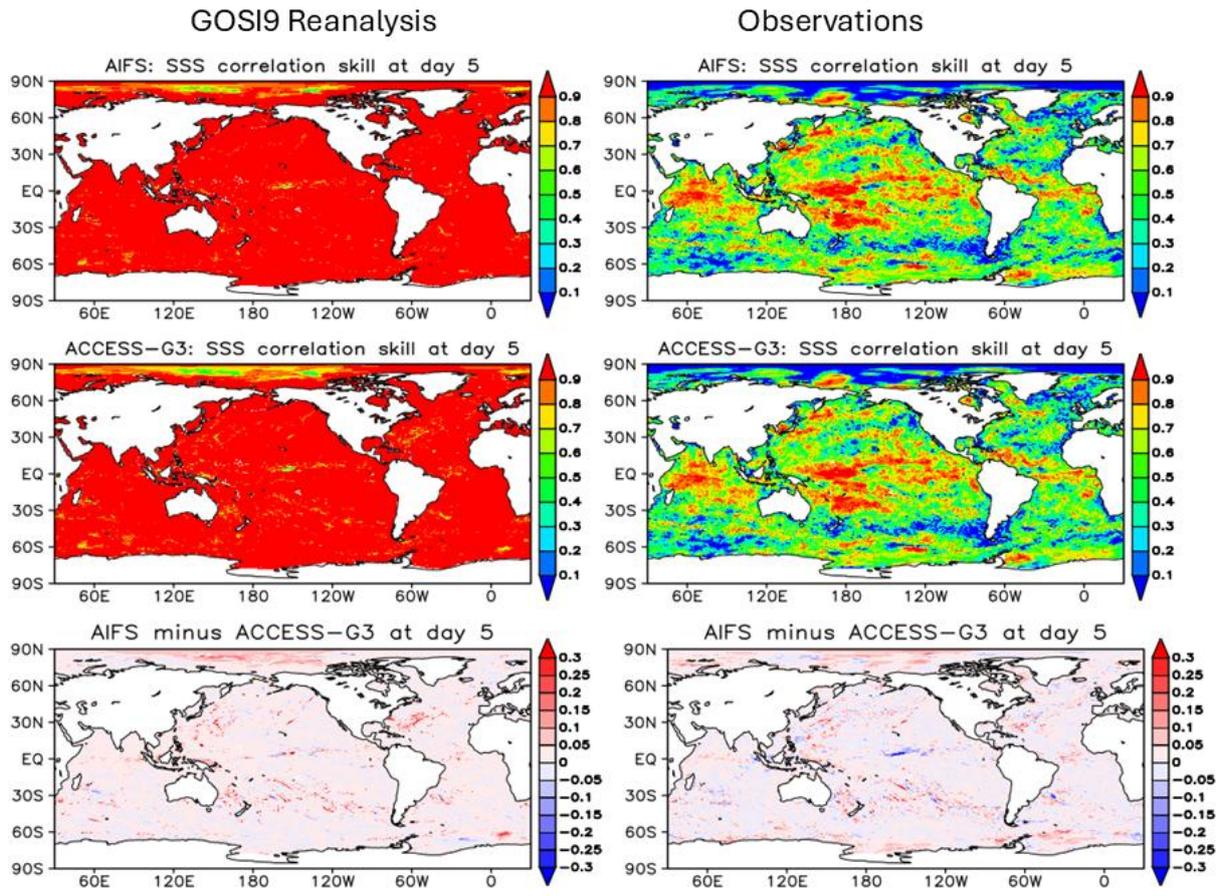

**Fig. S18.** Same as Fig. S17 but for day 5.

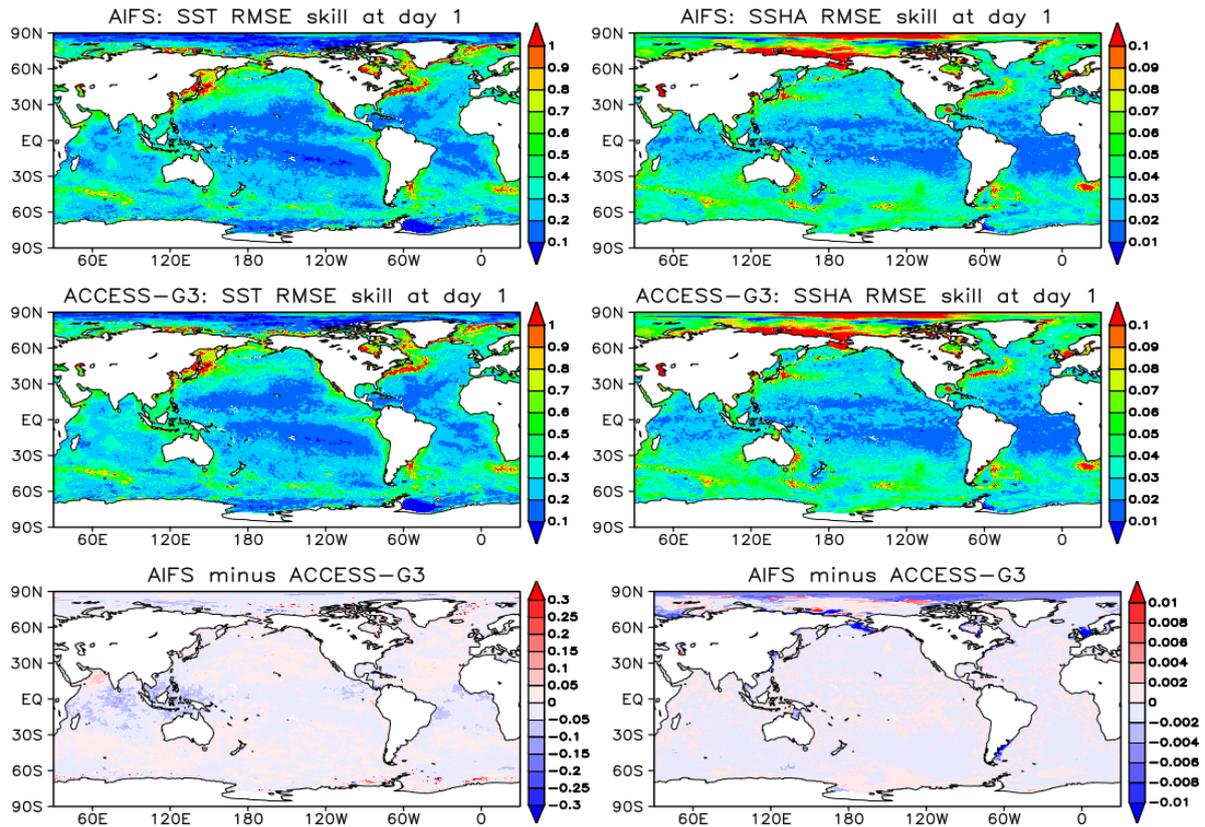

**Fig. S19.** RMSEs between ocean forecasts and observations at day 1 for SST (left column) and SSH (right column). The top row shows forecasts forced with AIFS data, the middle row shows forecasts forced with ACCESS-G3 data and the bottom row shows the correlation differences between the two. The units are °C for SST and m for SSHA.

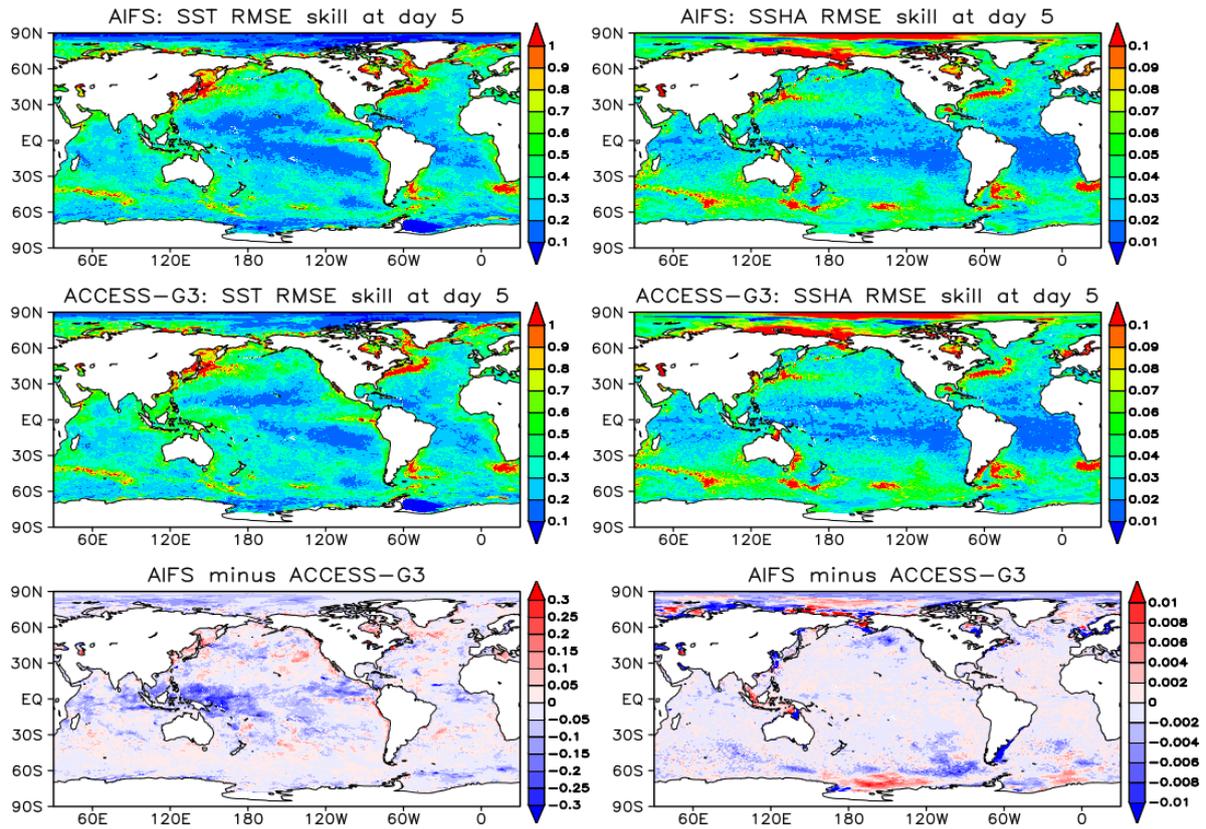

**Fig. S20.** Same as Fig. S19 but for day 5.

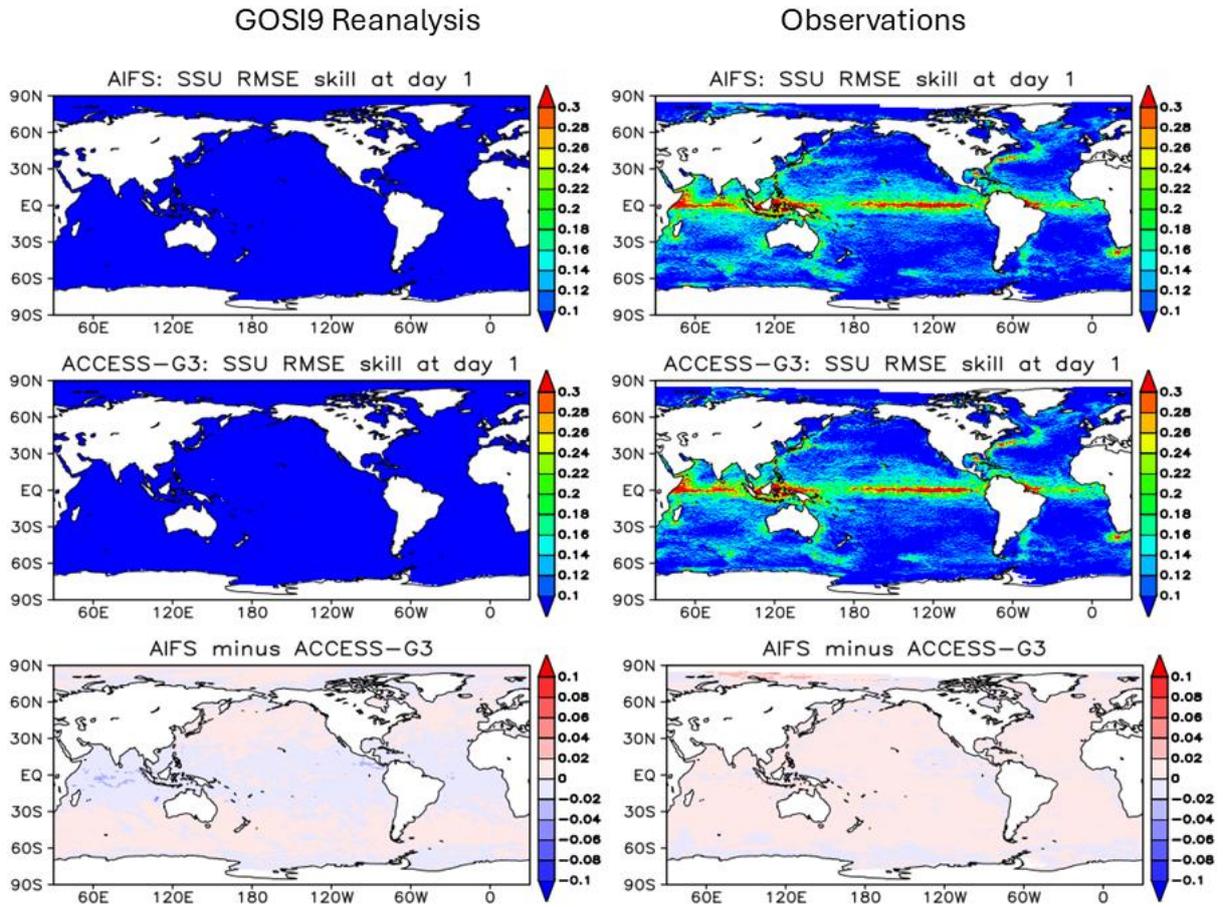

**Fig. S21.** RMSEs at day 1 between model-forecasted sea surface zonal current (SSU) and GOSI9 ocean reanalysis (left column), and between model SSU and observations (right column). The top row shows forecasts forced with AIFS data, the middle row shows forecasts forced with ACCESS-G3 data and the bottom row shows the correlation differences between the two. (unit: m/s)

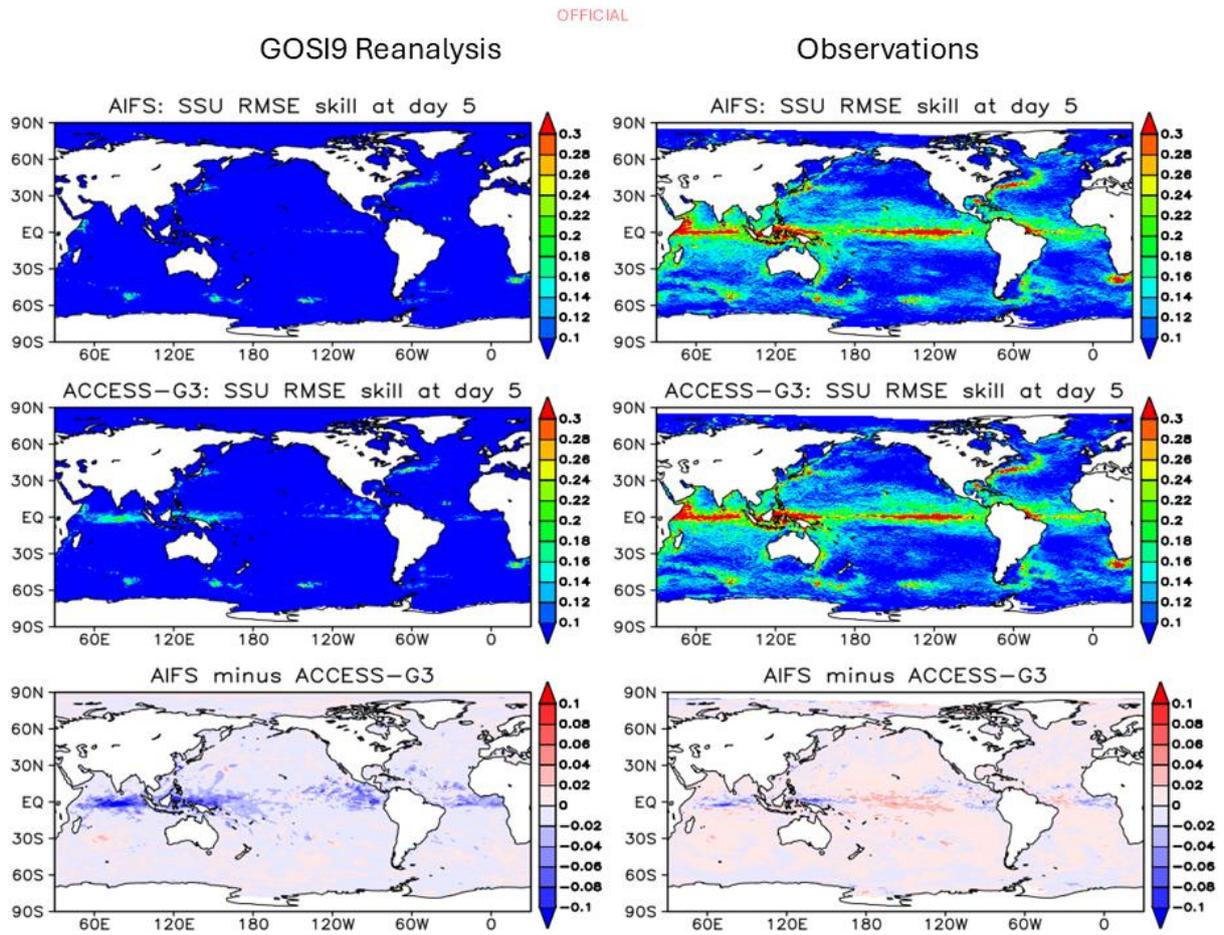

**Fig. S22.** Same as Fig. S21 but for day 5.

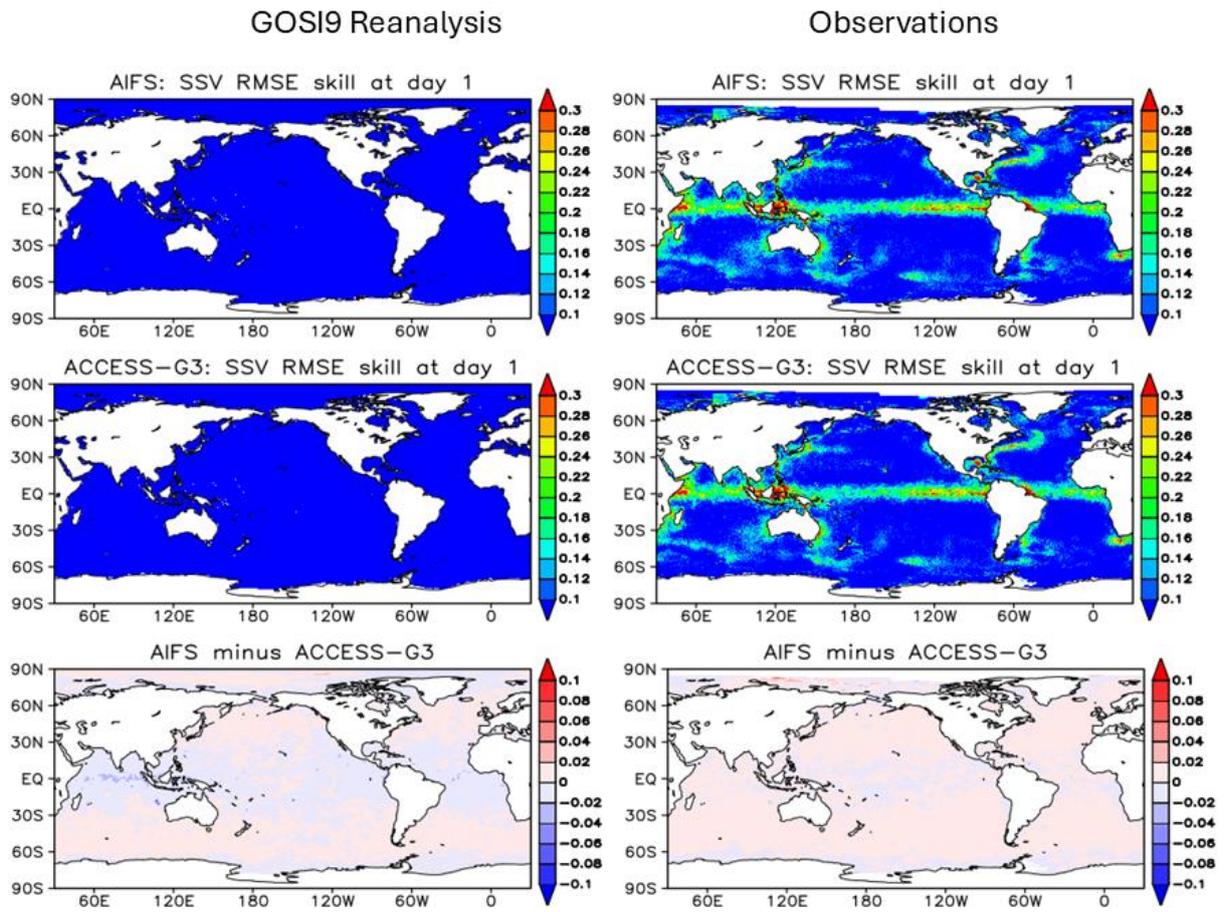

**Fig. S23.** Same as Fig. S21 but for sea surface meridional velocity (SSV).

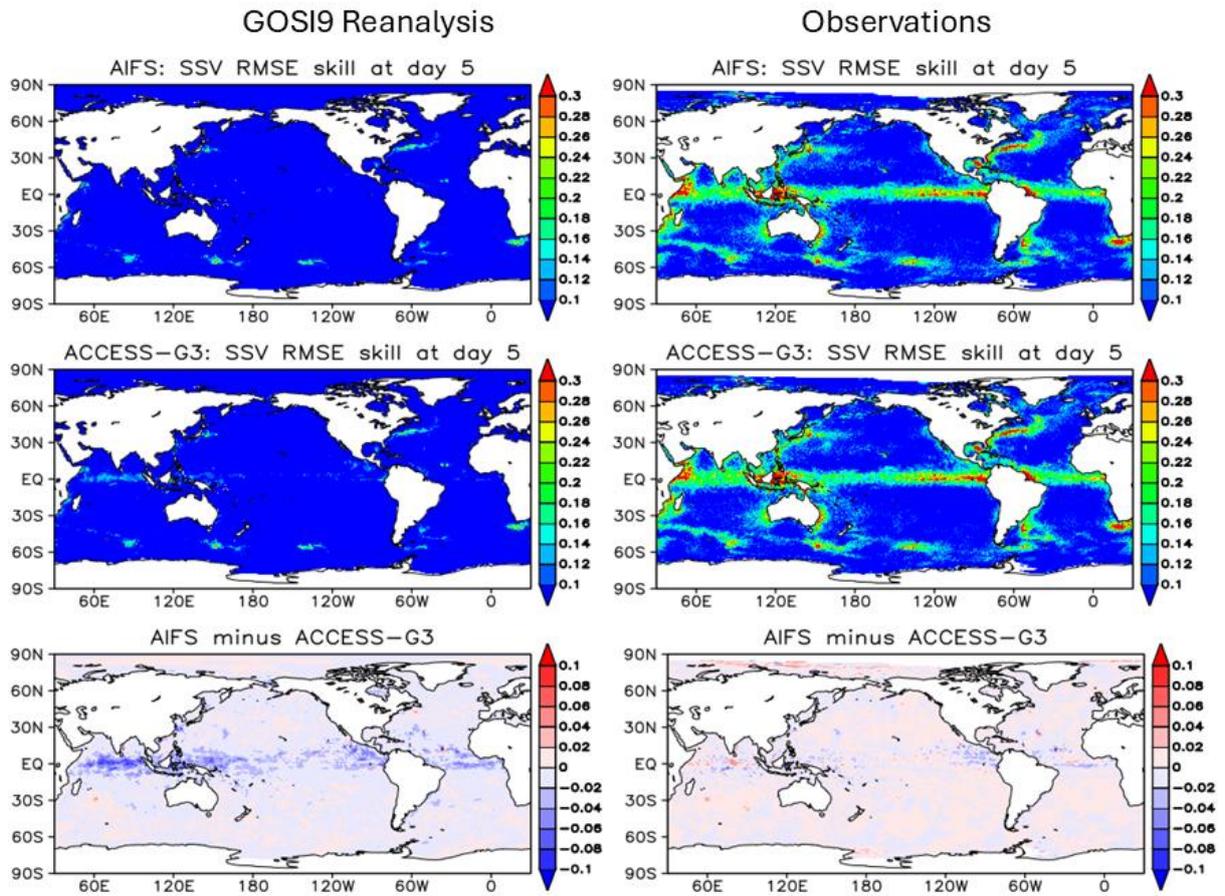

**Fig. S24.** Same as Fig. S23 but for day 5.

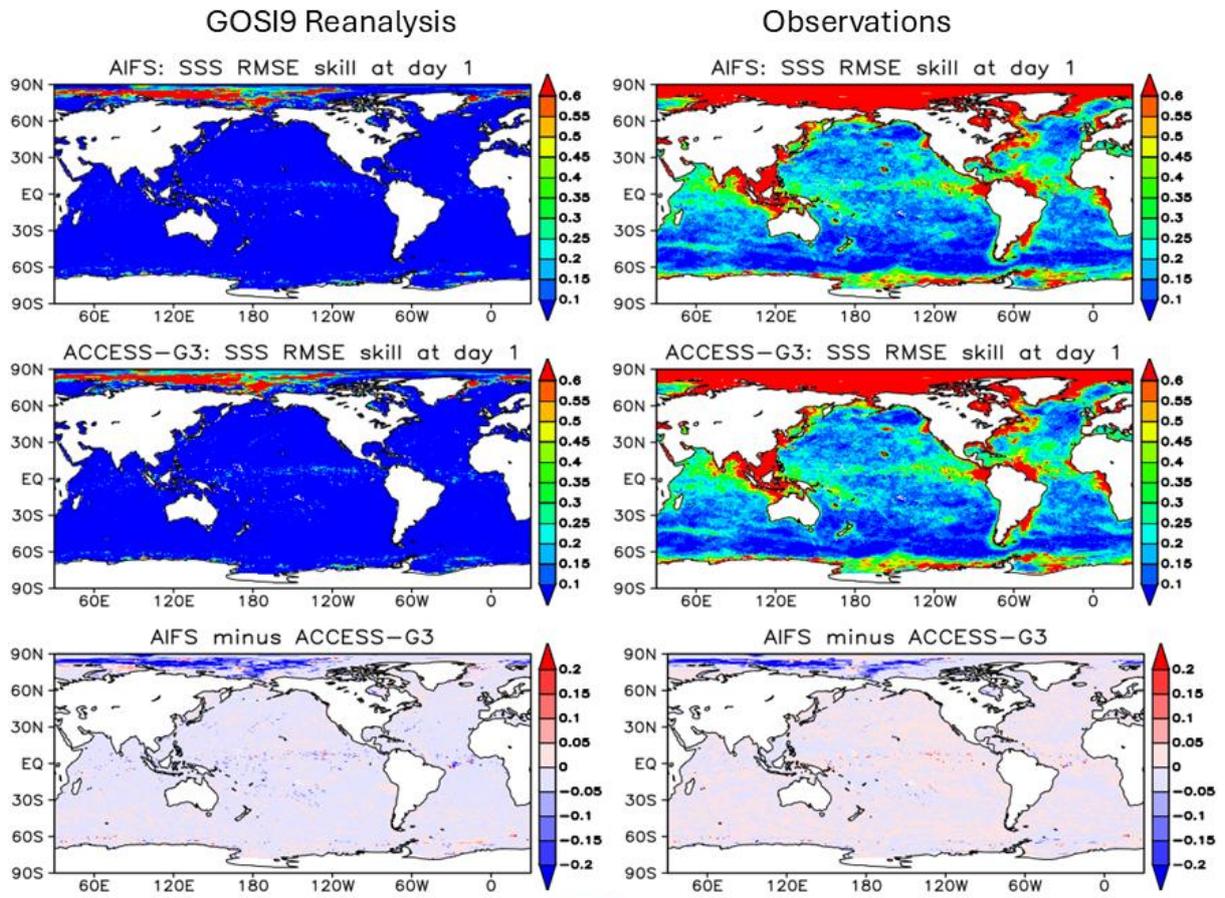

**Fig. S25.** Same as Fig. S21 but for sea surface salinity (SSS) (unit: psu).

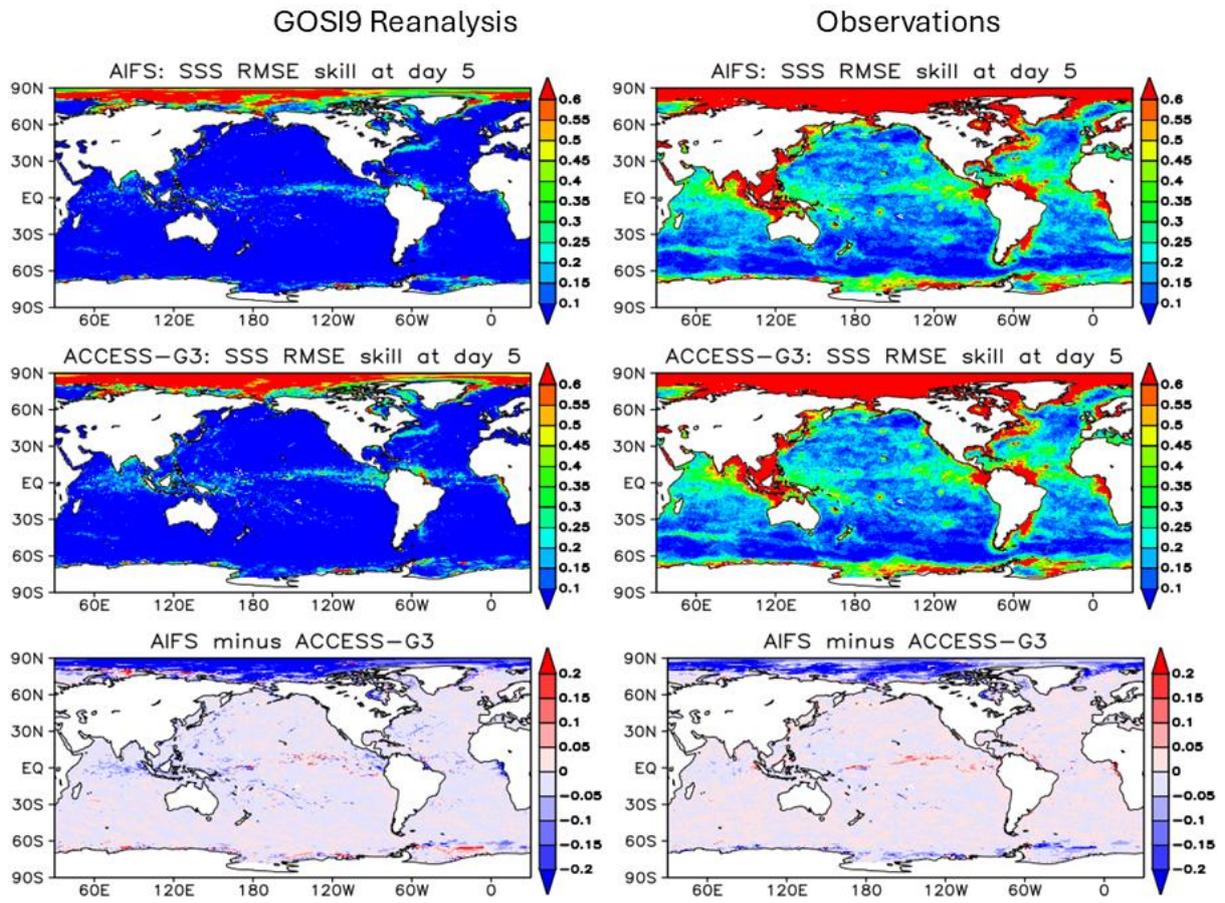

**Fig. S26.** Same as Fig. S25 but for day 5.